\documentclass[adp,fleqn]{w-art}
\usepackage{amsmath,amssymb}
\usepackage{times}
\usepackage{w-thm}

\theoremstyle{plain}

\theoremstyle{definition}

\usepackage[]{graphicx}
\chardef\bslash=`\\ 

\newcommand{\E}{{\rm e}}
\newcommand{\D}{{\rm d}}
\newcommand{\I}{{\rm i}}
\newcommand{\Tr}{{\rm Tr} }
\renewcommand{\Re}{{\rm Re}}
\renewcommand{\Im}{{\rm Im}}

\begin{document}


\DOIsuffix{theDOIsuffix}
\Volume{12}
\Issue{1}
\Copyrightissue{01}
\Month{07}
\Year{2003}
\pagespan{1}{}


\title[Charge transport in disordered interacting electron systems]
   {Quasiclassical theory of charge transport in disordered interacting electron systems}


\author[P.\ Schwab]{P.\ Schwab\footnote{Corresponding author\quad 
E-mail: {\sf Peter.Schwab@Physik.Uni-Augsburg.de}}\inst{1}}
\address[\inst{1}]{Institut f\"ur Physik, Universit\"at Augsburg, 86135 Augsburg, Germany}
\author[R.\ Raimondi]{R.\ Raimondi\footnote{E-mail: {\sf raimondi@fis.uniroma3.it}}\inst{2}}
\address[\inst{2}]{NEST-INFM and Dipartimento di Fisica, Universita di Roma Tre,\\ 
Via della Vasca Navale 84, 00146 Roma, Italy}
\begin{abstract}
We consider the corrections to the Boltzmann theory of electrical transport
arising from
the Coulomb interaction in disordered conductors.
In this article the theory is formulated in terms of quasiclassical Green's functions.
We demonstrate that the formalism is equivalent to the conventional diagrammatic
technique by deriving the well-known Altshuler-Aronov corrections to the conductivity.
Compared to the conventional approach, the quasiclassical theory has the advantage
of being closer to the Boltzmann theory, and also allows description of interaction
effects in the transport across interfaces, as well as non-equilibrium phenomena in the same theoretical framework.
As an example, by applying the Zaitsev boundary conditions which were originally developed
for superconductors, we obtain
the $P(E)$-theory of the Coulomb blockade
in tunnel junctions.
Furthermore we summarize recent results obtained for the non-equilibrium transport in thin films, wires and fully
coherent conductors.
\end{abstract}
\maketitle                   
\tableofcontents




%
%
\section{\label{Intro}Introduction}
\begin{figure}
\begin{center}
 \includegraphics[width=0.4\textwidth]{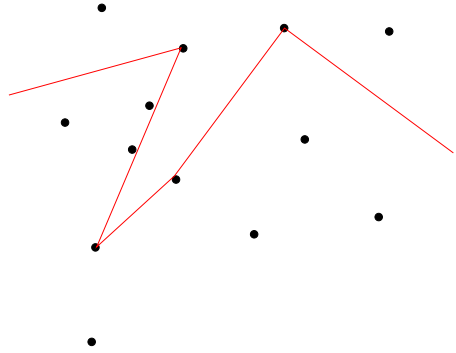} \\
\end{center}
\caption{The classical picture for charge transport in metals: electrons
move through the metal along classical trajectories. From time to time they scatter and transfer
momentum to the lattice.}
\label{figIntro1}
\end{figure}

More than a hundred years ago Drude put forth a theory of
metallic conduction \cite{drude00,ashcroftmermin}. In his theory electrons are treated as a gas of particles
which are assumed to move along classical trajectories until they collide with one another
or with the ions. The collisions abruptly alter the velocity of the electrons, as illustrated
in Fig.~\ref{figIntro1}.
Since that time several  details of Drude's theory have been improved.
In Drude's time, for example, it seemed 
reasonable to assume that the electronic velocity distribution at equilibrium  was given
by the Maxwell-Boltzmann distribution. About a quarter of a century later,
Sommerfeld replaced the Maxwell-Boltzmann distribution by the Fermi-Dirac distribution.
A further important step was the description of collisions beyond the relaxation time approximation,
which finally led to the Boltzmann equation for the dynamics of the distribution function.
In many cases the transport theory based on the Boltzmann equation
successfully describes the electrical conductivity of metals.
On the other hand the Boltzmann theory still assumes that electrons move along classical trajectories, and
quantum mechanical interference effects are neglected. The latter become important at a high
concentration of defects or in very small systems.
Therefore, in the limits of strong disorder or small system size,
the Boltzmann equation fails and the transport becomes non-classical.

\begin{figure}
\begin{center}
\includegraphics[width=0.8\textwidth]{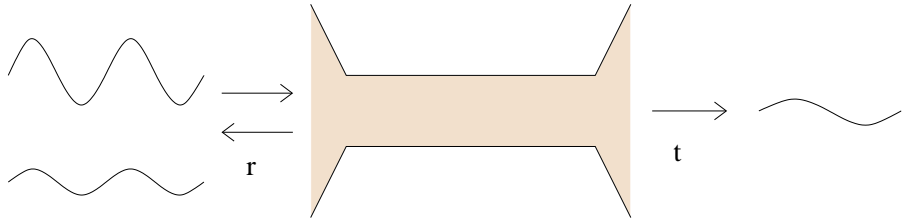}
\end{center}
\caption{The quantum mechanical picture for electron transport:
an incoming electron wave is transmitted or reflected.}
\label{figIntro2}
\end{figure}

Alternatively, following Landauer~\cite{landauer70}, one can adopt 
a quantum mechanical description of charge  transport:
Consider, for example, a constriction as shown in Fig.~\ref{figIntro2}.
An incoming electron wave is transmitted or reflected at the constriction with
amplitudes $t$ and $r$.
The current is then proportional to the transmission probability, {\em i.e.}~the squared modulus of
the amplitude.
For simplicity, we restrict ourselves to zero temperature and a one-dimensional conducting channel for now.
By assuming two different chemical potentials in the left and right reservoir, $\mu_L = \mu_R + eV$,
the current through the constriction is given by
\begin{equation}
I= 2 e \int_{\mu_R}^{\mu_L} \! \! \D \epsilon \, {\cal N}(\epsilon) v(\epsilon) |t|^2
;\end{equation}
the factor two is due to the electron spin. The product of the density of
states ${\cal N}(\epsilon)$ times the velocity $v(\epsilon)$
does not depend on energy and is equal to
$ {\cal N} v = 1/2\pi  \hbar$.
As a result the conductance of the constriction is determined as
\begin{equation}\label{eqIntro2}
G = 2 \frac{e^2}{ h } |t|^2
,\end{equation}
with $h/e^2\approx 25.8 \, {\rm k\Omega}$.

The connection between the scattering amplitudes and the conductance
was first proposed by Landauer~\cite{landauer70}.
Generalizations of the conductance formula (\ref{eqIntro2}), including for example
many transport channels and more than two leads, have been discussed intensively in the literature
\cite{economou81,fisher81,buttiker85,buttiker86,baranger89,meir92}.
The Landauer approach to the conductance has been successfully
applied to describe transport through
structures where the quantum mechanical coherence of the electron wave functions
is maintained over the full system.
A fascinating example is the
transport through quantum point contacts, which has been studied using both, semiconductor devices
\cite{vanwees88,wharam88}
as well as single atoms~\cite{scheer97}.

In the following,
we will focus on the transport in metallic systems, where the classical approach to the
conductivity is still a good starting point,
but quantum effects give rise to various types of corrections to the Drude conductivity.
The two corrections that will be considered in this review are weak localization, which
is a pure one-particle phenomenon,
and  effects arising from the combination of disorder and electron-electron  interactions.

In order to give a simple idea of how these quantum corrections arise and
to highlight the connection between the semiclassical  and Landauer's approach,
we present here a physical picture of the origin of weak localization.
For reviews on this subject see
\cite{altshuler85,bergmann84,lee85,chakravarty86,kramer1993,belitz94}.
The weak localization correction
to the conductivity arises as a result of the quantum interference of electron waves
in disordered systems.
\begin{figure}
\begin{center}
\includegraphics[width=0.5\textwidth]{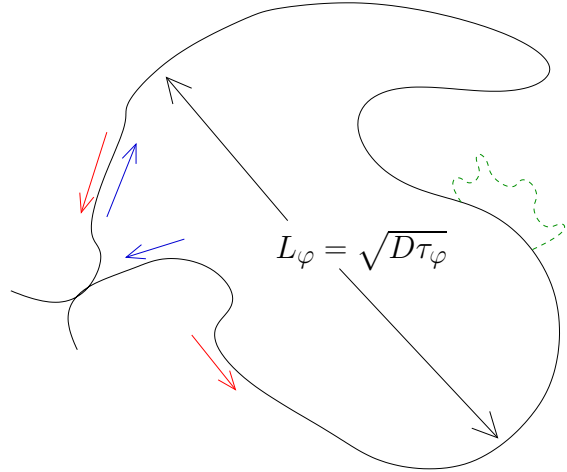}
\end{center}
\caption{A self-intersecting path as it is relevant for the weak localization correction to the conductivity.
The dashed wavy line represents an inelastic scattering event which is responsible for phase breaking; $\tau_\varphi$ is
the phase coherence time, and $D$ is the diffusion constant.}
\label{figIntro3}
\end{figure}
To travel from point $A$ to point $B$ electrons move along classical paths.
In order to obtain the total probability for a transfer from
point $A$ to $B$,
in classical physics,
one has
to sum the probabilities for a particle to move along all possible paths.
In quantum mechanics one has to sum the amplitudes of each path
and take the modulus squared at the end.
The relevant paths for weak localization are those with self-intersections and with
the velocities of the incoming and outgoing paths in opposite directions
as shown in Fig.~\ref{figIntro3}.
An electron can travel around the path clockwise or anti-clockwise.
The two paths can be assigned an amplitude $\Psi_1$ and $\Psi_2$;
the probability is then proportional to
\begin{equation}
|\Psi_1 + \Psi_2|^2 =  |\Psi_1|^2 +|\Psi_2|^2+ 2 \Re( \Psi_1^* \Psi_2 )
.\end{equation}
The first two terms on the right hand side correspond to the classical probabilities.
The third term, giving the interference between $\Psi_1$ and $\Psi_2$, only
appears in quantum mechanics.
For the special type of paths considered, the interference term is always positive, {\em i.e.}~due to interference
the probability of the process is enhanced.
Neglecting  the interference corresponds to a classical description of the electrons (Drude-Sommerfeld
theory, Boltzmann equation).

This article summarizes
recent contributions to the theoretical description of transport in mesoscopic
conductors.
The central questions addressed are the following:
\begin{itemize}
\item In which way is transport modified by electron-electron interactions?
\item What happens in the presence of a large voltage, when linear response theory fails?
\end{itemize}
Both questions are relevant for the interpretation of experiments:
Shortly after the discovery of weak localization, it was
found that similar effects in the conductivity are caused by the electron-electron interaction
\cite{altshuler78,altshuler79,altshuler80a,fukuyama80}.
Furthermore electron-electron interactions are also relevant
for weak localization itself, since they provide a mechanism for
phase breaking. However, while it is clear that inelastic scattering contributes to dephasing,
the exact way this happens is far less obvious, as is evident from the recent 
debates in the literature
\cite{golubev98,altshuler98,aleiner99,aleiner02a,raimondi99,cohen99,kirkpatrick02,volker02}.
 The importance of studying non-linear transport is also apparent:
Several transport experiments on mesoscopic samples are carried out at low temperatures.
Under these conditions inelastic scattering events which drive the electron system 
towards local equilibrium freeze out.
Therefore non-equilibrium conditions can be realized even at rather low voltages.

Here we will use the quasiclassical formalism which, 
as its historical predecessor, {\sl i.e}.\ the Boltzmann equation, is very powerful in
dealing with situations far from thermal equilibrium.
This was, for example, exploited in studying the dynamics of superconductors in the
1960s and 1970s \cite{redbook} and hybrid mesoscopic systems in the 1990s \cite{lambert98}.
In particular, we will show, that the quasiclassical method provides a convenient framework
to describe 
both quantum interference (weak localization) and
Coulomb interaction effects in normal-conducting systems out of equilibrium. 

\subsection*{Outline}
The article is arranged as follows:
In section \ref{chClassical} we introduce
the Green's function technique in
the non-equilibrium (Keldysh) formulation, which is the main theoretical tool 
used throughout this article. In particular, after giving the definition of the quasiclassical
Green's function, we derive the equation of motion as well as the boundary conditions
to be used in the presence of interfaces.
Following the literature we will demonstrate how a Boltzmann-like theory and the
Drude conductivity are recovered within this formalism.
In section \ref{chCoul} transport beyond the Drude-Boltzmann theory will be considered.
By extending the formalism of section \ref{chClassical} contributions to the current
 density due to
``maximally crossed diagrams'' and due to the Coulomb interaction will be calculated.

Sections \ref{chCoulApp} and \ref{SecNanNo} will be devoted to  applications, {\it i.e.},
the general expression for the current density of section \ref{chCoul} will be evaluated
explicitly for different experimental setups. To begin with, in section \ref{chCoulApp},
we will consider the limit
of weak electric field and derive the well-known expression for the Coulomb
correction to the Drude conductivity.
This will be followed by a discussion of the
the non-linear conductivity in films. Furthermore we will
investigate
the question of whether the phase coherence time $\tau_\varphi$, which is a central
quantity for weak localization,
is  relevant for the Coulomb interaction corrections to the conductivity as well.
In section \ref{SecNanNo} we will analyze non-linear transport in wires and in the presence
of interfaces. In order to illustrate the power of the method 
we first show how to obtain the Coulomb blockade theory.
We will also discuss the different regimes
that arise depending on how the length of the wire compares with the
thermal diffusion, inelastic electron-electron and electron-phonon lengths.
Chapter \ref{SecSummary} summarizes the main results.

A few technical details are provided in the appendices. In appendix
\ref{SecAppA} we outline the correspondence of the quasiclassical formalism
with the field-theoretic approach based on the non-linear sigma model.
In appendix \ref{SecAppB}, we provide the technical details necessary to include the spin degrees
of freedom.


\section{The classical theory of transport}
\label{chClassical}
The traditional transport theory in metals is based on the Boltzmann equation.
The central object is the distribution function $f({\bf k}, {\bf r}, t)$,
where $f({\bf k}, {\bf r},t )\D {\bf r} \,  \D {\bf  k} /(2\pi)^3$ is the number of electrons (per spin direction)
at time $t$ in the phase space volume $\D {\bf r} \, \D {\bf k}$.
The charge and current densities are given by
\begin{eqnarray}
\rho ({\bf r}, t) &=& (-2e) \int {\D {\bf k} \over (2\pi)^3} f({\bf k}, {\bf r}, t), \\
{\bf j} ({\bf r}, t) &=& (-2e) \int {\D {\bf k} \over (2\pi)^3}{ \hbar {\bf k} \over m} f({\bf k}, {\bf r}, t).
\end{eqnarray}
In thermal equilibrium the distribution function reduces to the Fermi function.
The Boltzmann equation determines the dynamics of the distribution function,
and reads
\begin{equation} \label{eqGreen19}
\partial_t f + {\bf v} \cdot \nabla_{\bf r} f + {1\over \hbar}{\bf F} \cdot 
\nabla_{\bf  k } f = I[f] 
.\end{equation}
The left hand side of the above equation
contains information on the energy spectrum, $ \hbar {\bf v } = \nabla_{ \bf k } \epsilon({\bf k })$,
and the external forces ${\bf  F}$, 
whereas the collision term on the right hand side describes the scattering processes.
For example, for impurity scattering, the collision term is given by 
\begin{equation} \label{eqGreen20}
I[f] = - \int { \D {\bf k }' \over ( 2  \pi )^3 } W_{{\bf k}, {\bf k}'}[f({\bf k}) - f({\bf k}')]
.\end{equation}
In this section we will recall how a Boltzmann-like kinetic 
equation is derived within the Green's function formalism.
To this end we use the
non-equilibrium Green's function technique, as originally formulated by Keldysh \cite{keldysh64}.
Our notation will mainly follow \cite{rammer86}.
After giving a few  general definitions,
we will  introduce the quasiclassical Green's function.
In the presence of impurity scattering, within the Born approximation
for the self-energy, a Boltzmann-like kinetic equation
and the Drude formula for the conductivity are
recovered. To go beyond the semiclassical treatment and
obtain the quantum corrections to the conductivity,
to be discussed in sections \ref{chCoul} -- \ref{SecNanNo},
further approximations are necessary.
From now on we set $\hbar = k_B =1$, except in
the final results, where, for clarity, we re-introduce the physical constants.

\subsection{The Keldysh formalism}
\begin{figure}
\centerline{\includegraphics[width=8cm]{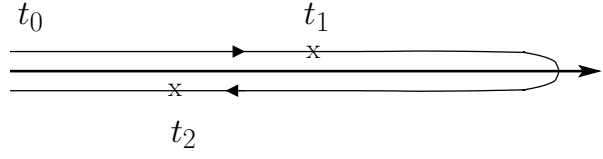}}
\caption{\label{figContour}A closed time path contour, passing through $t_1$ and $t_2$.}
\end{figure}

An object possessing a simple perturbation expansion is the so-called contour-ordered 
Green's function defined by
\begin{equation}
G(x_1, x_2) = - \I \langle T_c \Psi(x_1) \Psi^\dagger(x_2)  \rangle,
\end{equation} 
where $\Psi$ and $\Psi^\dagger$ are the usual fermion operators in the Heisenberg picture, 
and $x_i= ({\bf x}_i, t_i)$.
The brackets $\langle \dots \rangle$ denote an average over a statistical ensemble.
The times $t_i$ vary on a (complex) contour which starts at some time $t_0$,
and passes once through
$t_1$ and $t_2$ as shown in Fig.\ \ref{figContour}. The contour-ordering operator $T_c$ orders the Fermi
operators along the contour.
Here we consider a special contour, the ``Keldysh contour''. It consists of two parts: 
The first part starts at $t_0 = - \infty$ and ends at $t= +\infty$, whereas
the second part starts at $+ \infty$ and goes back to $-\infty$.
The contour-ordered Green's function on this contour can then be mapped on a two by two matrix 
in ``Keldysh space'',
\begin{equation}
G(x_1, x_2) \to \check G = \left(  \begin{array}{cc} G_{11} & G_{12} \\
                                                   G_{21} & G_{22} \end{array} \right) 
\end{equation}
with
\begin{eqnarray}
G_{11}(x_1, x_2) & = & - \I \langle T  \Psi(x_1) \Psi^\dagger(x_2)  \rangle  \\
G_{12}(x_1, x_2) & = & + \I \langle \Psi^\dagger(x_2) \Psi(x_1) \rangle \\
G_{21}(x_1, x_2) & = & -\I  \langle \Psi(x_1) \Psi^\dagger(x_2) \rangle \\
G_{22}(x_1, x_2) & = & - \I \langle  \tilde T  \Psi(x_1) \Psi^\dagger(x_2)  \rangle
;\end{eqnarray}
here $T$ and $\tilde T$ are the ordinary time-ordering and 
anti time-ordering operators.
It follows from the definition that the components of $\check G$ are not independent, and it is convenient to transform the matrix
$\check G$ as
\begin{equation}
\check G \to {1\over 2} 
\left( \begin{array}{rr} 1 & -1 \\ 1 & 1  \end{array}  \right)
\left( \begin{array}{rr} 1 & 0  \\ 0 & -1 \end{array}  \right)
\check G
\left( \begin{array}{rr} 1 & 1 \\ -1 & 1  \end{array} \right)
.\end{equation} 
In this new representation the 
Green's function is
\begin{equation} \label{eqGreen21}
\check G = \left( \begin{array}{cc}
G^R & G^K \\
G^Z  & G^A
\end{array} \right)
,\end{equation}
with
\begin{eqnarray} \label{eqGreen22}
G^R(x_1,x_2) & =  &-\I \Theta( t_1 -t_2)
 \langle \Psi(x_1) \Psi^\dagger(x_2)
 + 
\Psi^\dagger(x_2)\Psi(x_1)\rangle  \\
\label{eqGreen23}
G^A(x_1,x_2) & =  &+\I \Theta( t_2 -t_1)
\langle \Psi(x_1) \Psi^\dagger(x_2)
 + 
\Psi^\dagger(x_2)\Psi(x_1)\rangle   \\
G^K(x_1,x_2) & =  &-\I
\langle \Psi(x_1) \Psi^\dagger(x_2)
 - 
\Psi^\dagger(x_2)\Psi(x_1)\rangle  \\
G^Z(x_1, x_2) & = & 0 
.
\end{eqnarray}
To appreciate the physical meaning of the various Green's functions,
 let us consider first non-interacting electrons.
The retarded and advanced Green's functions can then be expressed in terms of the
eigenfunctions $\psi_\lambda$ and eigenenergies $\epsilon_\lambda$ of the single-particle Hamilton operator,
so that
\begin{equation}
G^{R(A)}_\epsilon({\bf x}_1, {\bf x}_2) = \sum_\lambda 
{\psi_\lambda({\bf x}_1) \psi_\lambda^*({\bf x}_2)\over \epsilon - \epsilon_\lambda \pm \I 0}
.\end{equation} 
We emphasize that  these two components of the matrix
Green's function depend only on the energy spectrum of the system, whereas
the Keldysh component of the Green's function carries the information about the
statistical occupation of the states.
In thermal equilibrium the Keldysh component
may be expressed in terms of the retarded and advanced components as
$G^K_\epsilon = [1-2 f(\epsilon)] (G^R_\epsilon-  G^A_\epsilon )$,
where $f(\epsilon)$ is the Fermi function.
Generally, the equation of motion for $G^K$ constitutes the
quantum-kinetic equation.
Approximations to this equation lead to the Boltzmann equation, Boltzmann-like
equations, and generalizations.

\begin{figure}
\noindent
\begin{center}
 \hspace{0.5cm}
\includegraphics[width=5.5cm]{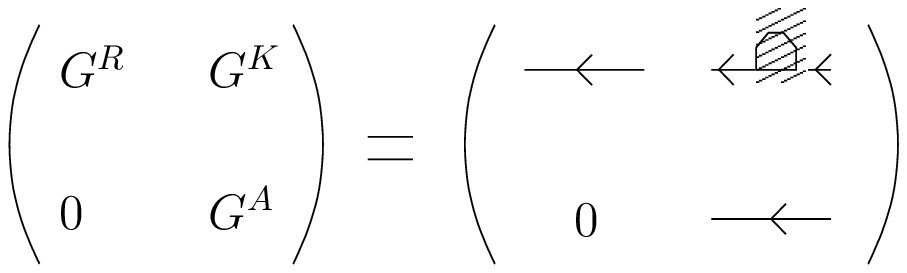}
\vspace{1cm}
\caption{
Graphical representation of the Green's function;
we mark $G^K$ with a shaded box, which can be understood as
the Keldysh component of the self-energy, see Eq.~(\ref{eqGreen27}).}
\label{fig1}
\end{center}
\end{figure}

In order to find such an equation we start with
the right-hand and left-hand Dyson equations which read
\begin{equation} \label{eqGreen25}
( \check G_0^{-1} - \check \Sigma ) \check G = \delta (x_1-x_2), \quad
\check G ( \check G_0^{-1} - \check \Sigma ) = \delta (x_1-x_2)
,\end{equation}
where the Green's functions $\check G_0$, $\check G$, and the self-energy $\check \Sigma$ 
are considered as matrices in space, time, and the Keldysh index.
$\check G_0$ is diagonal in the Keldysh index, the space and time dependence being given by
($e=|e|$)
\begin{equation}
\check G_0^{-1}(x_1, x_2)= 
\left[\I \partial_{t_1} -{1\over 2 m}\left(-\I \nabla_{{\bf x}_1}+ e {\bf A}(x_1) \right)^2 
+ e \phi(x_1) + \mu \right] \delta(x_1 - x_2 )
.\end{equation}
Disorder and interactions are contained in the self-energy.
In the Keldysh space the self-energy has the same triangular matrix structure as the Green's function, 
\begin{equation} \label{eqGreen26}
\check \Sigma = \left( \begin{array}{cc}
\Sigma^R & \Sigma^K \\
0   & \Sigma^A
\end{array} \right)
.\end{equation}
Relation (\ref{eqGreen26}) allows us to express
the Keldysh component of the Green's function as
\begin{equation} \label{eqGreen27}
 G^K = G^R \Sigma^K G^A
.\end{equation}
For a graphical representation see Fig.~\ref{fig1}.

\subsection{The quasiclassical Green's function}
In this subsection we show how to obtain a Boltzmann-like equation.
We start by
defining the center-of-mass and relative variables
\begin{equation}
\begin{array}{ll}
{\bf x}={1\over 2}({\bf x}_1 + {\bf x}_2), & {\bf r}= {\bf x}_1 - {\bf x}_2 \\[0.1cm]
\end{array}
\end{equation}
and Fourier transform with respect to the relative coordinate $\bf r $ in order to obtain the Green's function in
the mixed representation
\begin{equation}
G({\bf x}_1, {\bf x}_2 ) \to G({\bf p}, {\bf x} ) = \int \D {\bf r}
\E^{-\I {\bf p } \cdot {\bf r} } G({\bf x }+ {\bf r}/2, {\bf x} - {\bf r}/2 )
.\end{equation}
In the mixed representation, a matrix product of two objects $A({\bf x}_1, {\bf x}_2 ) $
and $B ({\bf x}_1, {\bf x}_2 )$, as it appears
for example in the Dyson equation, becomes \cite{eckern81}
\begin{equation}
( A B )({\bf p}, {\bf x} ) = \E^{-\I(\partial^A_{\bf x} \partial^B_{\bf p}  
              - \partial^A_{\bf p} \partial^B_{\bf x} )/2} A({\bf p}, {\bf x} ) B({\bf p}, {\bf x} )
.\end{equation} 
When both $A$ and $B$ are slowly varying in $\bf x$ a gradient expansion is justified,
{\em i.e}.\ the exponential in the equation above is expanded and only the first term or first few 
terms are kept.

By subtracting the right-hand from the left-hand Dyson equation and performing 
a gradient expansion, we obtain the equation of motion for the Green's function as
\begin{equation} \label{eqGreen28b}
\left[ \I \tilde \partial_t + \I{1\over m }  {\bf p } \cdot \tilde \partial_{\bf x} 
\right] \check G({\bf p}, {\bf x} ) =  \left[ \check \Sigma({\bf p},{\bf x}) , \check G({\bf p}, {\bf x}) \right]
.\end{equation}
Here the scalar potential $\phi$ and vector potential ${\bf A}$ are included
 in the gauge invariant
derivatives,
\begin{eqnarray} \tilde \partial_t \check G &= & 
\left[  \partial_{t_1} +  \partial_{t_2} - \I e \phi(t_1) + \I  e \phi(t_2) \right] \check G_{t_1, t_2 }({\bf p}, {\bf x})  \\
&=& \left[  \partial_{t}- \I e \phi(t+\eta/2) + \I  e \phi(t-\eta/2) \right] \check G({\bf p}, \eta;{\bf x}, t)
,\end{eqnarray}
and
\begin{equation}
\tilde \partial_{\bf x} \check G = \left[ \partial_{\bf x}+ \I e {\bf A}(t_1) - \I e {\bf A}(t_2) \right] \check G_{t_1 t_2}({\bf p}, {\bf x}) 
.\end{equation}
In the equation above, $t$ and $\eta$ are center-of-mass and the relative time, defined by
\begin{equation}
t={1\over 2}(t_1 + t_2),\quad  \eta = t_1 -t_2
.\end{equation}
We introduce next the $\xi$-integrated (quasiclassical)
Green's function
\begin{eqnarray} \label{eqGreen29}
\check g_{ {t_1} {t_2} } ( {\hat {\bf p} }, {\bf x} )
&= & {\I \over \pi} \int \D \xi
\D {\bf r} \E^{- \I {\bf p} \cdot {\bf r} }
\check G\left( {\bf x}+{ {\bf r} \over 2}, t_1;
{\bf x}-{ {\bf r} \over 2}, t_2\right)
 \\
&=& \check g( {\hat {\bf p} },\eta; {\bf x}, t ), \end{eqnarray}
where $\xi = {\bf p}^2/2m - \mu $, and ${\hat {\bf p}}$ is a unit vector
along the momentum.
The $\xi$-integration is understood as a principal value integration, which implies
that occasionally suitable (equilibrium) quantities have to be subtracted, in order
to ensure convergence.
In the entire article we will keep the notation of small $\check g$ for the $\xi$-integrated Green's
functions, and capital $\check G$ for the original Green's functions.
When approximating the density of states as an energy independent constant, the
$\xi$-integration is related to the original integration over the momentum ${\bf p}$
according to
\begin{equation} \label{eqGreen32}
\int {\D {\bf  p }\over (2\pi)^3 } \to
{ N}_0 \int \D \xi \int {\D \hat {\bf  p }\over 4 \pi}
.\end{equation}
Neglecting the $\xi$-dependence of the self-energy $\Sigma$, and integrating (\ref{eqGreen28b}) leads to the
equation of motion for the quasiclassical Green's function (Eilenberger equation \cite{eilenberger1968})
\begin{equation}
\label{eilenberger}
\left[  \tilde \partial_t + v_F  \hat{ \bf p } \cdot \tilde \partial_{\bf x}
\right] \check g(\hat{ \bf p}, {\bf x} ) = -\I \left[ \check \Sigma(\hat {\bf p},{\bf x}) , 
\check g( \hat {\bf p}, {\bf x}) \right].
\end{equation}
We note that the non-homogeneous term appearing on the right-hand side of the
Dyson equation has canceled when 
deriving the equation of motion for the quasiclassical Green's function. As a consequence,
the latter is determined only by a multiplicative constant.
Therefore the equation of motion has to be supplemented by a normalization condition
for the Green's function,
which turns out to be of the form ``$\check g^2 = 1$'', {\em i.e}.
\begin{equation}
\int \D t_1 \check g_{t t_1}(\hat{ \bf p}, {\bf x} ) \check g_{t_1 t'}(\hat{ \bf p}, {\bf x} )  = \delta(t-t') \check 1
,\end{equation}
where $\check 1$ denotes the
unit matrix in the Keldysh space \cite{eilenberger1968,shelankov1985}.
We now give some relations that are specific for impurity scattering,
when it is treated within the self-consistent Born approximation.
By assuming a Gaussian, $\delta$-correlated impurity potential
one observes that the impurity self-energy
is related to the $s$-wave part of the quasiclassical Green's function,
\begin{equation} \label{eqGreen33}
\check \Sigma_{t_1t_2}(\hat {\bf p}, {\bf x})  = -{\I \over 2 \tau}
\int {\D \hat {\bf p} \over 4 \pi} \check g_{t_1t_2} ({\hat {\bf p}} ,{\bf x})
,\end{equation}
where $\tau$ is the usual scattering time.
Due to the matrix structure of the theory, the above equation is satisfied by all 
 three components (retarded, advanced, Keldysh) of the self-energy.
The retarded and advanced self-energies have a simple structure,
$\Sigma^{R(A)}_{t_1t_2}=\mp ({\I /2 \tau}) \delta(t_1-t_2)$,
and the retarded and advanced Green's functions are given by
\begin{equation} \label{eqGreen34}
g^{R(A)}_{t_1 t_2}(\hat {\bf p}, {\bf x})  =  \pm \delta(t_1-t_2)
.\end{equation}
We already mentioned that the equation of motion for the Keldysh component of the Green's
function constitutes the kinetic equation.
By also transforming to the mixed representation for the frequency/time variables the kinetic equation reads
in the present case
\begin{equation}
\left[ \tilde \partial_t
 + v_F {\hat {\bf p}} \cdot \tilde \partial_{\bf x} \right]
 g^K({\hat {\bf p}} , \epsilon ; {\bf x}, t )
 =
\label{eqGreen37}
-\frac{1}{\tau}\left[ g^K({\hat {\bf p}},\epsilon; {\bf x}, t )-
  \int{\D {\hat{ \bf  p} }\over 4 \pi } g^K({\hat {\bf p}}, \epsilon ; {\bf x}, t ) \right]
 ,\end{equation}
which clearly reminds us of the Boltzmann equation.
A direct connection with the Boltzmann equation has been suggested in \cite{prange64}.
By defining the distribution function as
\begin{equation}
f( {\bf p } , {\bf x}, t ) = {1\over 2}\left[
1- {1 \over 2}g^K({\hat {\bf p}},\epsilon; {\bf x}, t )|_{\epsilon=\xi - e \phi }  \right]
,\end{equation}
and ignoring the explicit dependence of $g^K$ on $\hat{ \bf p}$ so that the derivatives of $f$ are
approximately given by
\begin{eqnarray}
\nabla_{\bf p } f & = & - {1\over 4 } \nabla_{\bf p} \xi \partial_\epsilon g^K \\
\nabla_{\bf x } f & = & - {1\over 4 }[ \nabla_{\bf x} g^K- e (\nabla_{\bf x} \phi )\partial_\epsilon g^K ]  \\
\partial_t  f & = & - {1\over 4 } [\partial_t g^K -e (\partial_t \phi ) \partial_\epsilon g^K]
,\end{eqnarray}
one recovers
from the kinetic equation (\ref{eqGreen37})
the Boltzmann equation (\ref{eqGreen19}) with the external force given by
${\bf F}= e(  \nabla_{\bf x} \phi + \partial_t {\bf A})$.

Besides the kinetic equation we also need a rule to calculate the physical observables.
Generally, the charge and current densities are related to a combination of all components of
the Green's function,
\begin{eqnarray} \label{eqGreen39}
\rho({\bf x}, t)   \! &=& \!
\I e \left[ G^K({\bf x}, t ; {\bf x}, t)- G^R({\bf x}, t ; {\bf x}, t)+G^A({\bf x}, t ; {\bf x}, t) \right] \\
\label{eqGreen40}
 {\bf j}({\bf x},t) \! &=& \!
 {e\over {2m}}
\left[ \nabla_{\bf x} \! - \! \nabla_{\bf x'} \! + \! 2\I e {\bf A}({\bf x},t ) \right]
\left[ G^K-G^R+G^A \right]  ({\bf x}, t; {\bf x'}, t) |_{{\bf x'}={\bf x}}.
\end{eqnarray}
In terms of the quasiclassical Green's functions, the charge and current densities
are \cite{rammer86}
\begin{eqnarray} \label{eqGreen41}
\rho({\bf x},t) &=& 2e { N}_0 \left[
{\pi \over 2} \int {\D \epsilon  \over 2 \pi }\int { \D \hat{\bf  p}\over 4 \pi} g^K(\hat {\bf p},\epsilon; {\bf x}, t)
- e \phi({\bf x}, t) \right] \\
\label{eqGreen42}
 {\bf j}({\bf x}, t) &=& e \pi { N}_0 \int {\D \epsilon \over 2 \pi }\int { \D \hat {\bf p}\over 4\pi}
  v_F \hat {\bf p}  g^K({\hat {\bf p}},\epsilon ; {\bf x},t ).
\end{eqnarray}

\subsection{Diffusive limit}
In this paper we will focus our discussion on the diffusive regime,
characterized by small external frequencies and momenta such that 
$\omega \tau, q v_F \tau \ll 1 $.
Under these conditions
the Green's function becomes almost isotropic and may be expanded
in spherical harmonics. 
In the present case, it is sufficient to retain the $s$- and $p$-wave components,
\begin{equation} \label{eqGreen43}
\check g(\hat {\bf p},{\bf x}) = \check g_s({\bf x}) + \hat {\bf p } 
\cdot  \check {\bf g}_p({\bf x})
.\end{equation}
By inserting Eq.~(\ref{eqGreen43}) into the kinetic equation
and separating the $s$- and $p$-wave parts, one obtains the relation
\begin{equation} \label{eq48a}
\check {\bf g}_p({\bf x}) = - v_F \tau \tilde \partial_{\bf x} \check  g_s({\bf x}),
\end{equation}
and finally the kinetic equation for the $s$-wave component of the Green's function 
\begin{equation}
\left(
\tilde \partial_t  -
 D \tilde \partial_{\bf x} \tilde \partial_{\bf x}  \right)
g^K_s(\epsilon;  {\bf x} , t) = 0
,\end{equation}
where we have introduced the diffusion constant  $D= v_F^2 \tau /3$.
From Eqs.\ (\ref{eqGreen41}) -- (\ref{eq48a}),
an explicit expression for the current density follows:
\begin{equation}
\label{eqGreen44}
{\bf j}({\bf x}, t)
=
-e \pi D  { N}_0  \tilde \partial_{\bf x}
\int {\D \epsilon \over 2 \pi}
g^K_s(\epsilon;{\bf x}, t)
=-D \nabla_{\bf x} \rho({\bf x},t)
+ 2 e^2 D { N}_0 {\bf E}({\bf x},t ),
\end{equation}
in agreement with the Drude formula for the conductivity.
\subsection{Boundary conditions}
\label{SecBoundary}
In particular, we will be interested in the transport properties of 
finite systems, {\em i.e}.\ metallic systems
with boundaries to insulators and reservoirs.
We note that, due to the $\xi$-integration and the gradient
expansion, 
the quasiclassical equations are not directly applicable across sharp interfaces
and therefore must be supplemented by boundary conditions.
For a surface with specular scattering, and for the non-interacting case we consider at present, 
the matching conditions read\footnote{Note that we use ``$T$'' for both, the transmission
probability and the temperature. The meaning will be apparent from the context.}
\begin{eqnarray} \label{eq51a}
\check g(\hat {\bf p}; {\bf x}_-) & = & R \check g ( {\tilde {\hat {\bf p} } }; {\bf x}_-)  
                                    + T \check g (\hat {\bf p} ; {\bf x}_+),\\ 
\label{eq52a}
\check g( {\tilde {\hat {\bf p} } } ;{\bf x}_+)& = &
                                      R \check g(\hat{{\bf p}} ; {\bf x}_+ ) +
                                      T \check g( {\tilde {\hat {\bf p} } } ; {\bf x}_-)
,\end{eqnarray} 
where $T$ and $R$ are the transmission and reflection probabilities 
at the surface. 
\begin{figure}
\noindent
\hspace{5.0cm}
\includegraphics[width=4cm]{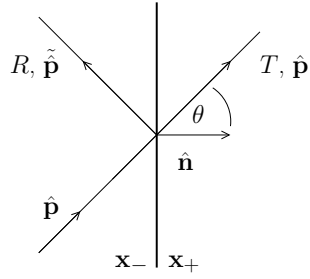}
\caption{\label{fig6a}An interface with specular scattering. An incoming particle moving in direction 
$\hat {\bf p}$ is reflected with probability 
$R$ into $\tilde{ \hat {\bf p }}$, and transmitted with probability $T$ into $ \hat{\bf p}$;
$\theta$ is the angle between the trajectory and the normal vector $\hat {\bf n}$ of the interface.
}
\end{figure}
$\hat {\bf p}$ and $\tilde {\hat {\bf p} }$ are the directions for incoming, transmitted, and reflected
particles as shown in Fig.\ \ref{fig6a}.
From the sum of the two equations and using the relation $R+T=1$ it is found that the antisymmetric combination
of $g(\hat {\bf p} )$ and $g( \tilde {\hat {\bf p} })$ is continuous at the interface,
\begin{equation}
\check g(\hat {\bf p}; {\bf x}_-) - \check g ( {\tilde {\hat {\bf p} } }; {\bf x}_-) = 
\check g(\hat {\bf p}; {\bf x}_+) - \check g( {\tilde {\hat {\bf p} } } ;{\bf x}_+)
,\end{equation}
from which one observes that
the current normal to the interface is conserved, since
\begin{equation}
\int{\D \hat {\bf p}\over 4 \pi  } ({\hat {\bf n}} \cdot { \hat {\bf p}}) \, \check g(\hat {\bf p}; {\bf x}_-) =
\int{\D \hat {\bf p}\over 4 \pi  } ({\hat {\bf n}} \cdot { \hat {\bf p}}) \, \check g(\hat {\bf p}; {\bf x}_+) 
.\end{equation}
After subtracting (\ref{eq52a}) from (\ref{eq51a}) 
and making use of the continuity of the antisymmetric combination
of the Green's functions it is seen that the symmetric combination
of the Green's functions jumps at the interface, with the size of the jump determined by
\begin{equation} \label{eq55a}
\check g(  \hat {\bf p}; {\bf x}_-) - \check g({ \tilde{\hat{\bf p} }}, {\bf x}_-) = {T \over 2 R} 
\left\{ 
\left[   \check g(\hat {\bf p}; {\bf x}_-) + \check g( { \tilde{\hat{\bf p} }}, {\bf x}_-) \right]
- \left[ \check g(\hat {\bf p}; {\bf x}_+) + \check g(\tilde{\hat{\bf p} },{\bf x}_+) \right]
\right\} 
.\end{equation}
The generalization of this boundary condition to the interacting case is given below, compare Eq.\ (\ref{eqZaitsev}).

After these general considerations we now focus our attention on the diffusive
limit. Equation (\ref{eq55a}) is valid both for clean and dirty metals.
In a dirty metal, however, a matching condition
which involves only the angular averaged Green's function $\check g_s({\bf x})$ instead of 
$\check g(\hat {\bf p};{\bf x})$ would be desirable. 
Recall that in the diffusive limit the angular dependence of $\check g( \hat {\bf p}, {\bf x})$ can be taken into account
by keeping only the $s$- and $p$-wave parts. 
One might then be tempted to expand the Green' functions in (\ref{eq55a}) in an $s$-wave and a $p$-wave part,
multiply (\ref{eq55a}) by $\hat {\bf p}$, take the angular average and hence obtain the matching conditions
in the diffusive limit in terms of the $s$-wave and $p$-wave components of the Green's functions.
However, the task of finding the matching condition turns out to be not that simple,
since close to the interface the angular dependence of the transmission and reflection
coefficients may generate higher harmonics in the Green's functions, and it is not justifiable to terminate 
the expansion after the first two terms.
Only at some distance from the interface will such an expansion hold.
The task is thus to derive an effective condition
which connects the Green's functions ``far'' from the interface, {\em i.e}.\ on distances of the order of the elastic
mean free path. 
By integrating the kinetic equation and making use of (\ref{eq55a})
it has been shown \cite{kuprianov88,laikhtman94,lambert97} that this matching condition
can be written as
\begin{equation} \label{EqMatchingCondition}
- D ( {\hat {\bf n}}\cdot  \tilde \partial_{\bf x} ) g_s^K({\bf x}) = \Gamma 
    \left[ g_s^K( {\bf x}_- ) - g_s^K( {\bf x}_+  ) \right]
,\end{equation}
where $\Gamma$ is a function of the angular dependent transmission and reflection amplitudes $R$ and $T$.
For a strong barrier ($T/R \ll 1$), $\Gamma$ is given by
\begin{equation}
\Gamma = v_F \int_{0}^{1}  \D(\cos\theta) \, \cos\theta \, {T( \cos \theta ) \over 2 R( \cos \theta ) },
\end{equation}
while for a weak barrier ($R/T \ll 1$)
\begin{equation}
\Gamma  = v_F  {1\over 9 } \left[ \int_0^1 \D(\cos\theta) \,  \cos^3(\theta ) {2R \over T}  \right]^{-1}
,\end{equation}
see \cite{lambert97};
for a barrier of arbitrary strength 
the parameter $\Gamma$ may be obtained from the solution of an integral equation \cite{lambert97}.

As a simple application of the matching condition (\ref{EqMatchingCondition})
we consider a system consisting of a diffusive wire of
length $L$ which is
attached to two reservoirs.
We study the system in a non-equilibrium
situation with an applied voltage $V_l-V_r=V$, where the subscripts
$l$ and $r$ indicate the left and right reservoirs, respectively.
The classical resistance of the structure is the  sum
of the wire resistance and the interface resistances,
$R_{\rm tot}=R_{\rm wire}+R_{l}+ R_r$,
and the current is $I=V/R_{\rm tot}$.
We will demonstrate now how this result can be reproduced within the quasiclassical formalism.

Within the quasiclassical formalism the current flowing in the wire is given by
\begin{eqnarray}
I_{\rm wire} &= &-e \pi D N_0 {\cal A} \int { \D  \epsilon \over 2 \pi} \partial_x g^K_s(\epsilon,x)
,\end{eqnarray}
where ${\cal A}$ the cross section, and
$x= 0 \dots L$ is the position along the wire.
The reservoirs are assumed to be in thermal equilibrium, with the distribution function given by 
$g^{K,l,r}_s(\epsilon) = 2 \tanh \left[(\epsilon + |e| V_{l,r})/2T \right]$.
The boundary conditions read
\begin{eqnarray} 
\label{eq60}
- D \partial_x g^K_s(\epsilon, x) |_{x=0} &=&  \Gamma_l [ g^{K,l}_s(\epsilon) - g^K_s(\epsilon, 0) ]  \\
\label{eq61}
- D \partial_x g^K_s(\epsilon,x)|_{x=L}   &=&  \Gamma_r [ g^K_s(\epsilon, L)- g^{K,r}_s(\epsilon) ]
,\end{eqnarray}
where $\Gamma_{l,r}$ are the transparencies of the left and right interface.
The current across the interfaces can be written as
\begin{eqnarray} 
I_l  & = &-{1 \over  2 } e \Gamma_l {\cal A} N_0 \int \D \epsilon [ g^K_s(\epsilon, 0) - g^{K,l}_s(\epsilon)] \\
I_r  & = &-{1 \over  2 } e \Gamma_r {\cal A} N_0 \int \D \epsilon [ g^{ K, r}_s(\epsilon) - g^K_s(\epsilon, L) ]
.\end{eqnarray}
Furthermore we assume that the wire is so short that we can neglect inelastic scattering.
In this case the kinetic equation for
the distribution function inside the wire simply becomes 
$-D\partial_x^2 g^K_s =0$.
Solving this equation
one finally finds the current as
$I= V/(R_l + R_{\rm wire}+ R_r)$, with
\begin{eqnarray}
R_r  &= & 1/( 2e^2 {\cal A} N_0 \Gamma_l ) \\
R_{\rm wire } &=&L/(2 e^2 {\cal A} N_0 D ) \\
R_l  &= & 1/( 2 e^2 {\cal A} N_0 \Gamma_r ) 
\end{eqnarray}
as one would expect for three resistors in series.


\section{Quantum interference in diffusive conductors -- general formalism} 
\label{chCoul}
Quantum effects give rise to deviations from the classical
expression of the electrical conductivity of a metal: 
The conductivity depends on the sample specific realization of the impurity potential,
and even after averaging over all possible realizations of the
impurity potential corrections to the conductivity remain.
The quantum corrections to the average conductivity
in a metal with diffusive electron motion
are weak localization, and the interaction contributions to the conductivity.
The latter are often classified as  
the particle-particle (Cooper) channel, which is
related to the exchange of superconducting fluctuations, and the particle-hole 
channel, which is related to the exchange of charge fluctuations (spin singlet channel) or 
spin fluctuations (spin triplet channel).
In this article we concentrate on the average conductivity, 
and in particular
on the Coulomb interaction in the particle-hole channel. 
We will neglect the Cooper channel. This is justified
in non-superconducting metals, since in this situation the relevant interaction parameter scales down under 
the renormalization group flow.
For completeness we will also discuss briefly weak localization. 
In this section we will give the general expressions for the contribution to the
current density due to weak localization and due to the electron-electron interaction.
The following sections \ref{chCoulApp} and \ref{SecNanNo} will contain specific applications.
%
%
\subsection{Weak localization} 
\begin{figure}
\begin{center}
\hspace{0.5cm}
\includegraphics[width=12cm]{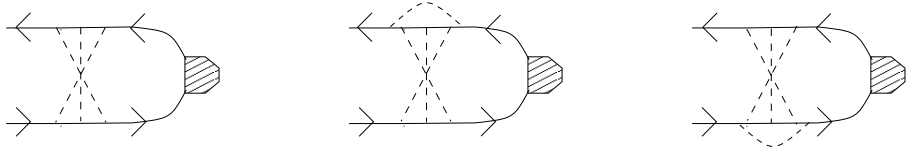}
\end{center}
\vspace{1cm}
\caption{Relevant graphs for the Keldysh component of the self-energy, ${\Sigma}^K_{\rm mc}$ 
(``mc'': maximally crossed). The
first graph leads to the term proportional to $g^K(- \hat{\bf p})$ in the kinetic equation,
the other two graphs are responsible for the term containing an angular average over $g^K(\hat{\bf p})$.}
\label{figCorr1}
\end{figure}

In a weakly disordered metal, quantum interference
leads at low temperature to deviations from the Drude-Boltzmann theory.
Gorkov {\em et al.}~\cite{gorkov79} and Abrahams {\em et al.}~\cite{abrahams79}
showed that the summation of maximally crossed diagrams
gives rise to divergences in the conductivity for arbitrarily weak disorder
in dimensions less than and equal to two.
This so-called weak localization correction to the conductivity is
due to electrons diffusing along closed paths, where quantum interference causes an enhanced
backscattering probability, as discussed in the introduction.
The weak localization contribution to the current density is given by
\begin{equation}\label{eqCorr1}
\delta {\bf j}_{\rm WL}(t) = -e^2 D  {4\over \pi}
\int_{\tau}^{\infty} \D \eta \, C^{t-\eta/2}_{\eta, -\eta}({\bf x},{\bf x})
{\bf E}(t-\eta)
,\end{equation}
where $D$ is the diffusion constant, the short time cut-off $\tau$ in the $\eta$-integration is the elastic scattering time,
and $C^{t}_{\eta, -\eta}({\bf x},{\bf x})$
is the cooperon at two coinciding points in space.
In the presence of a vector potential, the cooperon is given by the solution of the differential
equation
\begin{eqnarray} 
\left\{ 2{\partial \over \partial \eta} - D (\nabla_{\bf x}
+\I e {\bf A}_C )^2
\right\} C^{t}_{\eta ,\eta'}({\bf x},{\bf x}')
&& \nonumber\\
\label{eqCooperon}
=  \delta({\bf x} - {\bf x}' ) \delta( \eta- \eta' )
,\end{eqnarray}
with
${\bf A}_C = {\bf A}({\bf x},t+\eta/2)
+ {\bf A}({\bf x}, t-\eta/2 )$.
We recall now the results for
the conductivity.
In the absence of external fields the cooperon at two coinciding points in space is given by
\begin{equation}\label{eqCorr2}
C^t_{\eta, -\eta}({\bf x},{\bf x} ) = {1\over 2 }
\left( {1\over 4 \pi D \eta } \right)^{d/2} \E^{-\eta/\tau_\varphi }
,\end{equation}
where $d$ is the spatial dimension
and a phenomenological dephasing time $\tau_\varphi$ has been introduced;
microscopically dephasing arises from inelastic scattering.
Inserting the above into Eq.~(\ref{eqCorr1}), one arrives at
the standard result \cite{altshuler85,bergmann84,lee85,chakravarty86}
\begin{equation}\label{eqCorr3}
\delta\sigma_{\rm WL} = \left\{ \begin{array}{ll}
-{e^2 \over \pi \hbar }  \sqrt{ D \tau_\varphi }
& (d=1) \\[1ex]
- {e^2\over 2\pi^2  \hbar} \ln(\tau_\varphi / \tau )  & (d=2) \\[1ex]
{e^2 \over 2 \pi^2 \hbar}
\sqrt{1/ D \tau_\varphi} + {\rm const}& (d=3).
\end{array} \right.
\end{equation}
The connection to the Green's function formalism is the following:
In section \ref{chClassical} we demonstrated that, by approximating the impurity self-energy by
$\Sigma_{\rm Born}$, the Boltzmann equation for the distribution function and the Drude conductivity are
recovered. The kinetic equation including the weak localization correction
has been derived in \cite{hershfield86}, considering also the maximally crossed
diagrams,
\begin{equation}\label{eqCorr4}
\check \Sigma = \check \Sigma_{\rm Born} + \check \Sigma_{\rm mc}
,\end{equation}
with
\begin{eqnarray}
&&\includegraphics[height=0.65cm]{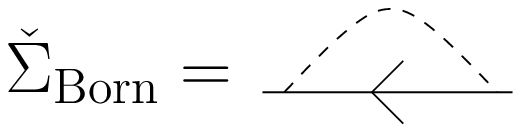} \label{eqCorr5} \\
&&\includegraphics[height=0.65cm]{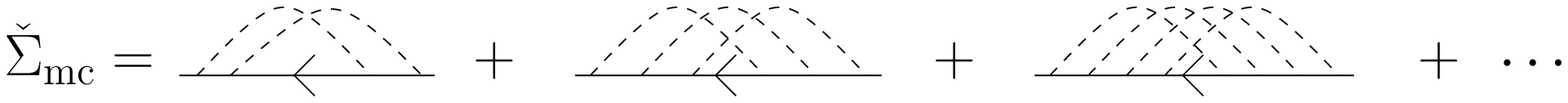} \label{eqCorr6}
\end{eqnarray}
The self-energy has to be determined self-consistently.
The contributions of the crossed diagrams to the retarded and the advanced self-energies
are small and are therefore neglected \cite{AGD65}.
The contribution to $\Sigma^K_{\rm mc}$ from the diagram with $m$ impurity lines contains
the sum of  products with ($2m-1$) Green's functions,
$\sum_{n=0}^{2m-2}( G^R )^n G^K ( G^A )^{2m-2 -n}$,
{\em i.e.}\
integrals involving all combinations
of the Green's function from
$G^K G^A \dots G^A$ to $G^R \dots G^R G^K$.
The momentum integrals reduce to integrals over pairs of Green's functions of the form
\begin{equation} \label{eqCorr9}
\eta^{RR}= {1 \over 2 \pi { N}_0 \tau }
\int {\D {\bf k} \over (2\pi)^3} G^R({\bf k},\epsilon) G^R(-{\bf k}+{\bf q},\epsilon-\omega )
,\end{equation}
and the analogously defined integrals $\eta^{AA}$ and $\eta^{RA}$.
For small ${\bf q}$ the integrals $\eta^{RR}$ and $\eta^{AA}$ are of the order $1/(\epsilon_F \tau)$, whereas the
integrals $\eta^{RA}$ are of order one,
\begin{equation}
\eta^{RA} \approx 1 - \tau [-\I \omega + Dq^2 ] 
.\end{equation}
Terms involving $\eta^{RR}$ or $\eta^{AA}$ will thus be neglected.
Following this rule the graphs contributing to
${\check \Sigma}_{\rm mc}$ are shown in Fig.~\ref{figCorr1}.
In order to understand how the above diagrams modify the kinetic equation, let us examine
the first diagram in Fig.~\ref{figCorr1}. In this diagram there is an equal number of retarded and
advanced Green's functions and a Keldysh Green's function in the middle, {\em i.e}., we have a sequence
$G^R ...G^RG^KG^A ... G^A$. 
The self-energy of the first diagram of Fig.~\ref{figCorr1}
has the form
\begin{equation}
\label{eqGreen76c}
\delta \Sigma^K ({\bf p},\epsilon ;{\bf q}, \omega )=
{1\over 2 \pi N_0 \tau^2}\sum_{{\bf Q}} C ({\bf Q}, \omega ) G^K({\bf Q}- {\bf p}, {\epsilon}; {\bf q}, {\omega} ),
\end{equation}
where Fourier variables ${\bf p},\epsilon$ and ${\bf q}, \omega$ refer to relative
and center-of-mass coordinates, respectively, and the integration over the cooperon
takes care of all the series of maximally crossed diagrams:
\begin{eqnarray}
C({\bf Q}, \omega ) & = &\tau  \left[  \eta^{RA} +  ( \eta^{RA} )^2 + \dots \right]= \tau \eta^{RA}/[1- \eta^{RA}]  \\
  &\approx & {1\over -\I \omega + D Q^2}
.\end{eqnarray}
To finally derive the modified
Eilenberger equation, the  
commutator between the self-energy and Green's function, which appears 
on the right-hand side of Eq.~(\ref{eqGreen28b}), has to be integrated 
with respect to $\xi$. 
As a result we have
\begin{equation}
\label{eqGreen76d}
 \frac{1}{\pi}\int \D\xi \, [{\delta \check \Sigma}, {\check G}]^K=
 \frac{1}{\pi}
{1 \over 2 \pi N_0 \tau^ 2 }\sum_{{\bf Q}} C ({\bf Q}, \omega ) 
\int \D \xi 
G^K 
(G^R -G^A)
.\end{equation}
Following \cite{hershfield86} we use the Ansatz
\begin{equation}
G^K({\bf p }, {\bf q} ) ={1\over 2} g^K({\hat {\bf p} }, {\bf q} ) \left(  G^R({\bf p}+ {{\bf q}/ 2 } ) -
G^A({\bf p}- {{\bf q}/ 2 } ) \right)
,\end{equation}
which is based both on the fact that $G^R$ and $G^A$ are not corrected beyond
the Born approximation, and that the equation above gives the 
correct quasiclassical Green's function upon $\xi$-integration.
A similar
analysis can be carried out for the other diagrams of Fig.~\ref{figCorr1}; 
after some algebra the kinetic equation is found as
\begin{eqnarray}
\left[ \tilde \partial_t + v_F \hat {\bf p}\cdot \tilde\partial_{\bf x} \right] g^K(\hat {\bf p},\eta ; {\bf x}, t ) & = &
-{1\over \tau}\left[g^K(\hat {\bf p}, \eta; {\bf x}, t)-
\int { \D \hat {\bf p}\over 4 \pi}g^K(\hat {\bf p}, \eta; {\bf x}, t) \right] \cr
\label{eqEilenbergerWL}
&&  \hspace{-2cm} +\frac{1}{\tau}\int_{-\infty}^{t} \D t'\alpha(t,t')
\left[g^K(-\hat {\bf p},  \eta; {\bf x}, t')- \int {\D \hat {\bf p}\over 4 \pi}g^K(\hat {\bf p}, \eta; {\bf x}, t' ) \right]  \\
\alpha(t, t') &=  &{2\over \pi N_0} C^{t-\eta'/2}_{\eta+\eta', \eta-\eta'}({\bf x}, {\bf x}), \quad \eta' = t-t'
,\end{eqnarray}
where we put the general version of the cooperon $C^{t}_{\eta, \eta'}({\bf x}, {\bf x})$ with three
different time arguments as given in (\ref{eqCooperon}).
The maximally crossed diagrams give rise to an additional 
scattering term on the right hand side of Eq.~(\ref{eqEilenbergerWL}) 
which
is non-local in time, with the dephasing time as the relevant time scale.

We restrict ourselfes now to the diffusive limit and calculate the change in the current density due to the maximally
crossed diagrams. One identifies two contributions. The first term appears because the additional scattering term
may change the solution of the kinetic equation for the $s$-wave  part of the Green's function,
$g^K_s \to g^K_s + \delta g^K_s$, and
\begin{eqnarray}\label{eqCorr12}
\delta {\bf j}_{\rm Born }({\bf x}, t) & = & - e \pi D {\cal N}_0 \tilde \partial_{\bf x}
\int {\D \epsilon \over 2 \pi}
\delta g^K_s( \epsilon; {\bf x}, t) \\
&=& - D \nabla_{\bf x} \delta \rho({\bf x}, t)
.\end{eqnarray}
The second term arises from the
modified relation between the $s$-wave and $p$-wave contribution to the Green's function,
\begin{equation}
{\bf g}_{p} = - v_F \tau  \tilde \partial_{\bf x} g_s +
v_F \tau \int_{-\infty}^t \D t' \alpha(t,t') \tilde\partial_{\bf x} g_s
\end{equation}
which leads to
\begin{eqnarray} \label{eqCorr10}
\delta {\bf j}_{\rm mc}({\bf x}, t) &= &
2 e D \tau \int_\tau^{\infty} \D \eta C^{t-\eta/2}_{\eta, -\eta}({\bf x},{\bf x})
\tilde \partial_{\bf x}\int {\D \epsilon \over 2 \pi }
      g^K_s(\epsilon;{\bf x},t - \eta )  \\
&=&  -{2 D \over \pi N_0} \int_{\tau}^\infty \D \eta C^{t-\eta/2}_{\eta, -\eta }({\bf x},{\bf x})
 \left[ - \nabla_{\bf x} \rho({\bf x}, t-\eta)  + 2e^2  N_0  {\bf E}({\bf x}, t-\eta) \right]
.\end{eqnarray}
The total weak localization correction to the current density is thus
\begin{equation}\label{eqCorr11}
\delta {\bf j}_{\rm WL } = \delta {\bf j}_{\rm Born} + \delta {\bf j}_{\rm mc}.
\end{equation}
In this equation the sum of both terms is needed in order to ensure charge conservation.
In the special case
where the electric field and the charge density are homogeneous in space,
the weak localization correction to the current density, as given in
Eq.~(\ref{eqCorr1}), is recovered.
%
\subsection{Interaction correction to diffusive transport}
Shortly after the discovery of weak localization it was found \cite{altshuler79,altshuler80a,fukuyama80}
that similar effects in the
conductivity are also caused by the electron-electron interaction.
The interaction correction to the conductivity in the particle-hole singlet channel, for example, is given
by \cite{raimondi99}
\begin{equation}\label{eqCorr13}
\delta \sigma_{\rm EEI} = - {4 e^2 D \over \pi d}
\int_\tau^\infty  \D \eta
\left({\pi T \eta \over  \sinh( \pi T \eta) } \right)^2
\left( {1 \over 4 \pi D \eta } \right)^{d/2}
\end{equation}
which leads to
\begin{equation}\label{eqCorr14}
\delta \sigma_{\rm EEI} \approx
\left\{ \begin{array}{ll}
-4.91 {e^2 \over \pi^2 \hbar } \sqrt{\hbar D/k_B T}      & (d=1) \\[1ex]
-{e^2 \over 2 \pi^2 \hbar} \ln( \hbar/k_B T \tau ) & (d=2) \\[1ex]
1.83 {e^2 \over 6 \pi^2 \hbar} \sqrt{k_B T/ \hbar D }  +{\rm const}.  & (d=3)
\end{array}
\right.
. \end{equation}
The inclusion of the triplet channels does not change the
functional form of the
temperature dependence of the correction to the conductivity,
but modifies the prefactor, which then depends also on the strength of the electron-electron interaction in the
spin triplet channel \cite{finkelstein83,castellani84}, see also Eq.~(\ref{eqSigmaTriplet}) below.

\begin{figure}
\begin{center}
\includegraphics[height=6cm]{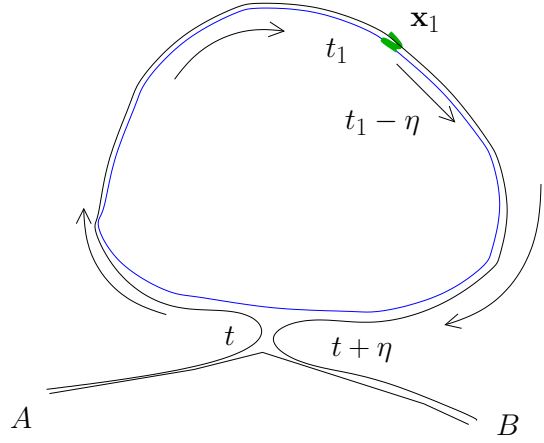}
\end{center}
\caption{Graphical representation of processes leading to the interaction correction to the conductivity.
The figure shows two paths for the propagation from $A$ to $B$. One of the paths contains an extra
closed loop. The corresponding phase factor $\exp(\I \varphi_{\rm loop})$ cancels with a phase factor due
to a virtual particle which is created at $({\bf x}_1, t_{1}-\eta)$, propagates
along the same closed loop, and is finally absorbed at $({\bf x}_1, t_1)$.}
\label{figCorr2}
\end{figure}

Whereas a simple and convincing physical interpretation of weak localization exists, we are not
aware of as simple an interpretation of the interaction effect.
However, attempts have been made \cite{bergmann87,rudin97}, and we present the main ideas.
First, one observes that the impurities perturb the charge distribution in the metal,
$n \to n + \delta n$.
For example, a single impurity gives rise to the so-called Friedel
oscillations of the density, which persist even at large distance
$\delta n (r) \sim \sin( 2 k_F r)/r^3$.
In the presence of many impurities the electron density becomes non-uniform, with the details
depending on the particular distribution of the impurity positions.
When both electron interaction and disorder are present,  the charge inhomogeneity
acts as an additional scattering potential, which,
within the Hartree approximation, is given by
\begin{equation}\label{eqCorr15}
V_H({\bf r}) = \int V({\bf r}- {\bf r}')\delta n({\bf r}') \D {\bf r}'
.\end{equation}
It is clear that this additional scattering potential may affect the elastic mean free path.
The way this happens, however, is less obvious.
In the following we give an argument \cite{bergmann87} which shows
that the interaction contribution to the conductivity is due to quantum interference.
Let us consider two classical paths an electron can follow to travel from point A to point B.
Let path ``1'' and ``2'' be identical up to an extra closed loop in path ``2'', so
that there is a phase difference between the two amplitudes,
$\Psi_2 = \exp (\I \varphi_{\rm loop}) \Psi_1$.
The sign of the interference term  $\Re( \Psi_1^* \Psi_2)$ is positive or negative, depending on
the phase $\varphi_{\rm loop}$,
and therefore processes of this type, when averaged over all the impurity
configurations, give a negligible small contribution to the total
probability for traveling from A to B.
This will be different in the presence of electron-electron interactions.
To first order in the Hartree potential $V_H$  we may write the interference term as
$\Re(\Psi_1^* \Psi_2 V_H )$.
Now note that
the charge inhomogeneity
$\delta n(\bf r)$, and therefore $V_H({\bf r})$, is related to (virtual) electrons or holes 
which propagate along
closed paths, since
\begin{equation}\label{eqCorr16}
\delta n({\bf r}) ={\I \over \pi } \int \D \epsilon f(\epsilon)
    \left[ G^R_\epsilon({\bf r}, {\bf r}) - G^A_\epsilon({\bf r}, {\bf r}) \right]
.\end{equation}
In the case where a virtual hole
goes around the same closed loop
as the path number ``2'',
the phase factor $\exp(\I \varphi_{\rm loop})$ cancels, and $\Psi_1$ and $\Psi_2$ interfere coherently.
A graphical representation of the relevant process
is shown in Fig.~\ref{figCorr2}.

The Coulomb correction to the electrical conductivity was investigated in the past
with various methods, ranging from standard many-body diagrammatic calculations of the Kubo formula
to a field-theoretic description based on a mapping of the original problem of interacting
Fermions to  a non-linear sigma model for matrix fields, and
also to a generalized kinetic equation approach
\cite{altshuler78,finkelstein83,castellani84,altshuler85,rammer86,strinati91,belitz94,kamenev99,chamon99,zala01}.
In the following we will briefly outline how the
Coulomb interaction in a disordered medium can be incorporated in the quasiclassical
formalism we employ in this article.
We start from the kinetic equation and the expression for the current density in the diffusive limit,
\begin{equation}
\tilde \partial_t \check g_s - D \tilde \partial_{\bf x} (\check g_s \tilde \partial_{\bf x}  \check g_s ) =  0 
\end{equation}
and
\begin{equation}
{\bf j}({\bf x},t)=-e\pi D N_0 ( \check g_s \tilde \partial_{\bf x} \check g_s )^K \!({\bf x}, t)
.\end{equation}
These two equations generalize the expressions we gave in section \ref{chClassical} 
to the situation where the quasiclassical retarded and advanced
Green's functions are not necessarily equal to plus or minus one.
The Coulomb interaction is now introduced by adding internal electric fields to the external one,
\begin{equation}
\phi \to \phi^{\rm ext} \check 1 + \check \phi^{\rm int}
,\end{equation} 
and physical observables are obtained after averaging over the internal fields.
Formally this can be carried out by means of a Hubbard-Stratonovich transformation
within a functional integral formalism. More physically, this approach corresponds to
describing the Coulomb interaction in terms of the exchange of scalar and longitudinal
photons. 
The internal field has a non-trivial structure in Keldysh space,
\begin{equation}
\check \phi^{\rm int} = \left(\begin{array}{cc} \phi_1  & \phi_2 \\
                                     \phi_2  & \phi_1  \end{array} \right)
,\end{equation}
and the field fluctuations are related to the retarded, advanced and Keldysh components of the
(screened) Coulomb interaction,
\begin{equation}
-\I e^2 \left( \begin{array}{cc} 
 \langle \phi_1 \phi_1 \rangle  &  \langle \phi_1 \phi_2 \rangle  \\
\langle \phi_2 \phi_1 \rangle &  \langle \phi_2 \phi_2 \rangle 
\end{array}\right) =  {1\over 2}
\left( \begin{array}{cc}
V^K & V^R \\
V^A & 0
\end{array}\right)
.\end{equation}
In this language, screening effects appear
as self-energy corrections of the photon propagators.
To stress the difference between an external classical field and an internal quantum
one, we observe that $\phi_1$ and $\phi_2$ are related to the sum and difference
of the values that the field takes in the upper and lower parts of the Keldysh
contour, respectively.  An external field takes the same value on both parts
of the Keldysh contour and hence its ($12$)- and ($21$)-components in 
Keldysh space vanish.
Finally we define the distribution function $F$ via
the $s$-wave part of the quasiclassical Green's function
according to
\begin{equation}
g^K_{s}  =  g^R_{s} F
               - F g^A_{s}
.\end{equation}
We now show how to
calculate quantities like the Green' functions, the density of states or
the current density perturbatively in the interaction.
For the Green's function, for example, we write
\begin{equation} \label{eq81a}
\check g([\phi^{\rm int}]) =
\left(\begin{array}{cc}
1 & 2F_0 \\
0 & -1      \end{array}  \right)
+
\left(\begin{array}{cc}
\delta g^R & \delta  g^K \\
\delta g^Z & \delta g^A
    \end{array}  \right)
.\end{equation}
Note that due to  $\phi_2\neq 0$,
the ($21$)-component, {\em i.e}.\ $\delta g^Z$ is nonzero. Causality is restored only after
averaging over the internal field fluctuations, and the ($21$)-component then vanishes. The normalization
condition for the Green's function, $\check g_s \check g_s  = \check 1$,
allows us now to express the variation of the
retarded and advanced Green's function and its average over the internal field as
\begin{eqnarray}
\delta g^R & \approx &  -{1\over 2 } g^K \delta g^Z; \quad
\langle \delta g^R \rangle_{\phi^{\rm int}} \approx
 -{1\over 2 }\langle  \delta g^K \delta g^Z \rangle_{\phi^{\rm int}}, \\
\delta g^A & \approx &  \hphantom{-}{1\over 2} \delta g^Z g^K; \quad
\langle \delta g^A \rangle_{\phi^{\rm int}}
\approx \hphantom{-}{1\over 2}\langle \delta g^Z \delta g^K \rangle_{\phi^{\rm int}}
.\end{eqnarray}
In order to determine the correction to the retarded Green's function
to first order in the screened Coulomb interaction
it is sufficient to determine $\delta g^K$ and $\delta g^Z$ to first order in the internal field.
The product of $\delta g^K$ and $\delta g^Z$ also enters the current density, since
\begin{eqnarray}
{\bf j} & = & - e \pi D N_0 \langle (\check g_s \tilde \partial_{\bf x} \check g_s )^K 
\rangle_{\phi^{\rm int}} \\
& \approx &
 - e \pi D N_0 \Big[   \tilde \partial_{\bf x} \left( 2 F_0 +  \langle  \delta g^K  
 \rangle_{\phi^{\rm int} } \right) \cr
&&
 + \langle  \delta g^K ( \tilde \partial_{\bf x} \delta g^Z )\rangle_{\phi^{\rm int}}
 F_0 + F_0 \langle ( \tilde \partial_{\bf x} \delta g^Z ) \delta g^K \rangle_{\rm \phi^{\rm int}}
 \Big]\label{103}
.\end{eqnarray}
To first order in the field $\check \phi^{\rm int}$ one finds from the equation of motion
\begin{eqnarray}
\left(\tilde \partial_t + D \tilde \partial_{\bf x} \tilde \partial_{\bf x} \right)  \delta g^Z(\eta;{\bf x}, t)
& = & 2 \I \phi_2({\bf x},t) \delta(\eta)  \\
\left(\tilde \partial_t - D \tilde \partial_{\bf x} \tilde \partial_{\bf x} \right)  \delta g^K(\eta;{\bf x}, t)
 & = & 2 \I [\phi_1 , F_0 ] + \dots 
\end{eqnarray}
where the time derivative $\tilde \partial_t$ contains the external field $\phi^{\rm ext}$,
but not the internal field $\check \phi^{\rm int}$.
The ``$\dots$'' indicate a term proportional to $\phi_2$; this term is of no
importance here, since the correlator
$\langle \phi_2 \phi_2  \rangle$ is zero. The times $\eta$ and $t$ are the relative and the 
center-of-mass times.
In order to solve the above equations it is useful to define the diffusion propagator (the diffuson)
via the following equation:
\begin{equation}
\label{eqCoul22}
\left\{ \tilde \partial_t - D \tilde \partial_{\bf x} \tilde \partial_{\bf x}
 \right\} D^{\eta}({\bf x}, t;{\bf x}', t')
  = \delta({\bf x} - {\bf x}' ) \delta( t-t' )
,\end{equation}
which allows us to express the product of $\delta g^K$ and $\delta g^Z$ as 
\begin{eqnarray} \label{eq105}
\int \D t' \langle \delta g^K_{t t'}({\bf x}) \delta g^Z_{t' t_1 }({\bf x}_1) \rangle_{\phi^{\rm int}} & = & 
 \langle \delta g^K_{t t_1}({\bf x}) \delta g^Z_{t_1 t_1 }({\bf x}_1) \rangle_{\phi^{\rm int}} \\
 & =  &
2 \I \int  \D {\bf x}' \D {\bf x}'' \D t' \D t''  
D^{t-t_1}\left({\bf x}, {t+t_1 \over 2 } ; {\bf x}', t'- {t-t_1 \over  2} \right) \nonumber\\
 && \times  F_0\left( t-t_1; {\bf x}', t'-{t- t_1\over 2 } \right)V^R({{\bf x}', t'; {\bf x}'', t''}) \nonumber\\
 &&  \times  D^{\eta  = 0}({\bf x}'', t''; {\bf x}_1, t_1). 
\label{eq106}
\end{eqnarray}
As an immediate application of this equation one may obtain the correction to the density of states in equilibrium.
The translational invariance both in time and space allows simplification of the above multiple integrals, and one
arrives at
\begin{equation}
\frac{ \delta N(\epsilon)}{N_0} = \Re\langle \delta g^R(\epsilon) \rangle_{\phi^{\rm int}}  = 
\Re \left[ - \I  
\sum_{\bf q}
 \int {\D \omega \over 2 \pi} 
\tanh \left({\epsilon - \omega \over 2 T } \right) \left( {1 \over -\I \omega + D q^2} \right)^2 V^R({\bf q}, \omega ) \right]
.\end{equation}

By collecting the various pieces, we find for 
the current density in the presence of Coulomb interaction 
the result
$\delta {\bf  j}_{\rm EEI} = \delta {\bf j}_{\rm Born} + \delta {\bf j}_V $,
with
\begin{eqnarray}
\label{eqCorr45}
\delta {\bf j}_{\rm Born} ({\bf x}, t) &=& - e \pi  D { N}_0  \nabla_{\bf x} 
\int  { \D  \epsilon \over 2 \pi } \langle \delta g^K(\epsilon; {\bf x}, t) \rangle_{\phi^{\rm int}} \\
\delta {\bf j}_{V}({\bf x}, t ) &=&
e 4 \pi  D { N}_0 
\int \D \eta \, \D {\bf x}' \,  \D t' \, \D {\bf  x}'' \,  \D t''\nonumber\\
&&\times \Re \big[
F_0(-\eta;{\bf x}, t-\eta/2) D^\eta({\bf x},t-\eta/2;{\bf x}', t' -\eta/2  )
F_0(\eta;{\bf x}', t'-\eta/2)
\nonumber\\
&& \times V^R({\bf x}',t';{\bf x}'', t'')
(-\I \nabla_{\bf x}) D^0({\bf x}'', t''; {\bf x}, t-\eta )
\big].
\label{eqCorr46}\end{eqnarray}
Notice that this expression of the current density is
valid for an arbitrary form of the distribution function
and of the diffuson. This allows us to investigate  electrical transport  in different
experimental and geometrical setups as will be carried out in the following two sections.

The expression for the current density (\ref{eqCorr46}) 
has been derived first in \cite{schwab01}, both using diagrammatic
techniques as well as with the 
Keldysh version \cite{kamenev99,chamon99,feigelman00}
of the non-linear sigma model \cite{finkelstein83}.
Equation (\ref{eqCorr46}) generalizes earlier results, which are valid in the absence of 
the external vector
potential \cite{nagaev94} or near local equilibrium \cite{raimondi99,leadbeater00}.
In \cite{leadbeater00} the spin-triplet channels and Fermi liquid renormalizations were included,
which allows to apply the theory even in strongly interacting systems.

The starting point of the derivation given here, the diffusion equation (\ref{eq81a}) in the
presence of internal fields, corresponds to the saddle point of the non-linear 
sigma model approach of \cite{kamenev99}. 
Details of the connection between the non-linear sigma model and the 
quasiclassical formalism are also given in the appendix.
Finally we mention that in \cite{zala01} a similar method was used to derive the kinetic equation in the
presence of interaction and disorder beyond the diffusive limit.
%
\section{The Coulomb interaction in diffusive conductors -- applications } 
\label{chCoulApp}
This section and the following are devoted to specific applications of the formalism.
We will outline how to recover
the well known Coulomb interaction correction to the linear conductivity within the present formalism.
This will be followed by considerations on phase breaking and gauge invariance.
\subsection{Linear conductivity} 
\label{SecLin}
To compute the 
linear conductivity
we apply an external electric field ${\bf E}$, by choosing 
${\bf A}=-t {\bf E}$, $\phi =0$. Furthermore we assume that the
electron distribution function has the equilibrium form, 
$F(\epsilon, {\bf x} )= \tanh(\epsilon/2T)$. Equivalently, in the time domain this means
$F(\eta;{\bf x},t) = - \I T /\sinh( \pi T \eta)  $.
For a system which is 
homogeneous in space 
(after averaging over the disorder),
the charge density is homogeneous, too, and therefore
$\delta {\bf j}_{\rm Born}$  vanishes.

In order to calculate $\delta {\bf j}_V$ the screened Coulomb interaction and
the diffuson are required.
The dynamically screened Coulomb interaction as a function of frequency and momentum reads 
\begin{equation} \label{eqAppCoul}
V^R({\bf q}, \omega) = \frac{4\pi e^2}{q^2+{8\pi e^2}{ N}_0 
\frac{D q^2 }{-\I \omega + Dq^2}  }
\approx \frac{1}{2{ N}_0 } \frac{-\I \omega + Dq^2}{Dq^2}
.\end{equation}
The expression in the middle of this equation is valid in three dimensions.
On the right hand side we have assumed good screening, {\it i.e.}\ the screening vector $\kappa$, with 
$\kappa^2= 8 \pi e^2 { N}_0$, is assumed to be large.
The perfectly screened Coulomb interaction in one and two dimensions is identical
to the one in three dimensions, as given on the right hand side of Eq.~(\ref{eqAppCoul}).
Next we evaluate the two diffusons $D^{\eta}({\bf q}, t, t')$ entering the current density. 
It is important to note that they 
appear with different time arguments $\eta$.
In the second of the two diffusons in $\delta {\bf j}_V$ the relative time $\eta$ is zero with
the consequence that the diffuson does not depend on the vector potential ${\bf A}$, and is
thus given by
\begin{equation}
D^0({\bf q}, \omega) =  \frac{1}{-\I \omega + Dq^2}
.\end{equation}
The convolution of the interaction with the second of the two diffusons appearing in the
formula for the current then gives 
\begin{equation}
\label{eqCorr47}
\int \D t_2 V^R({\bf q}, t_1 -t_2 ) D^0({\bf q}, t_2, t - \eta) = {1\over 2 { N}_0 }
{1\over D q^2}\delta(t_1 -t+ \eta )
,\end{equation}
and the expression for the current density becomes
\begin{equation}
\label{eqCorr48}
\delta {\bf j}_V(t) = - {2 e  \over  \pi}
\sum_{\bf q} \int_\tau^\infty \D \eta {{\bf q} \over q^2 }
\left({\pi T \over  \sinh (\pi T \eta ) } \right)^2
D^\eta({\bf q}, t- \eta/2, t-3\eta/2).
\end{equation}
The electric field enters via the remaining diffuson, which is given by
\begin{eqnarray}
\label{eqCorr49}
D^\eta({\bf q}, t -\eta/2, t-3 \eta/2) &= &  \exp[-D({\bf q}- e {\bf E}\eta)^2\eta ] \\
\label{eqCorr50}  & = & \E^{-Dq^2 \eta}
  \left(1 + 2D e {\bf q}\cdot {\bf E} \eta^2 + \dots \right)
.\end{eqnarray}
After performing the momentum integration one finally arrives at 
\begin{equation} 
\label{eqCorr49a}
\delta {\bf j}_V = - {4 e^2 D \over \pi d}
\int_\tau^\infty  \D \eta
\left({\pi T \eta \over  \sinh( \pi T \eta) } \right)^2
\left( {1 \over 4 \pi D \eta } \right)^{d/2}
{\bf E}
,\end{equation}
in full agreement with
equation (\ref{eqCorr13}) for the correction
to the conductivity. 

In the appendix it is shown how to include the electron-electron interaction in the spin triplet channel.
The correction to the current density from the spin triplet interaction is \cite{leadbeater00}
\begin{equation}
\delta {\bf j}_{\rm triplet} = - {2 e^2 D \over \pi }
A_0^a \left({1-A_0^a \over 4 \pi D}  \right)^{d/2}
\int_\tau^\infty { \D \eta \over \eta }
\left({\pi T \eta \over  \sinh( \pi T \eta) } \right)^2
\int_0^\eta \D t {t \over (\eta - A_0^a t )^{1+d/2}}
\, {\bf E}
,\end{equation}
from which the correction to the conductivity in one, two and three dimensions are found to be 
\begin{equation} \label{eqSigmaTriplet}
\delta \sigma_{\rm EEI} = \left\{ \begin{array}{ll}
-4.91 \left[ 1  - 3  \frac{2(1-A_0^a)^{1/2}-2 + A_0^a}{  A_0^a }  \right] 
{e^2 \over \pi^2 \hbar } \sqrt{\hbar D / k_B T } & d=1 \\[2mm]
- \left[ 1 + 3 \left(1+ \frac{1-A_0^a}{A_0^a} \ln( 1 - A_0^a)   \right)  \right] \frac{e^2}{2 \pi^2 \hbar}
\ln(\hbar/k_B T \tau )  & d=2 \\[2mm]
 1.83 \left[ 1 + 3 \frac{2(1-A_0^a)^{3/2}- 2 + 3 A_0^a }{A_0^a} \right] 
\frac{e^2 }{6 \pi^2 \hbar} \sqrt{k_B T / \hbar D}
& d=3
\end{array} \right.     
, \end{equation}
where $A_0^a$ is the standard Landau parameter for interaction in the spin triplet channel.
%
%
\subsection{Charge diffusion and induced electrical potential} 
\label{SecInducedPotential}
One might query the fact that the correction to the conductivity due to the Coulomb interaction
seems to be independent of the interaction strength, 
compare (\ref{eqCorr13}) or (\ref{eqCorr49a}).
Recalling the derivation, we note that this fact is related to the assumption of perfect screening.

In this section we will present a different method \cite{ingold92a} to calculate the screened Coulomb interaction,
or, to be more precise, to calculate the product of the screened interaction with the diffuson.
This clarifies further the validity of Eq.\ (\ref{eqCorr47}) in relation to perfect screening,
and has proven to be useful also in situations
where charging effects play a role.
First notice that the diffuson describes the spreading charge cloud $\rho_0({\bf x}, t)$
of a charge which is injected into the system at $({\bf x}_0, t_0)$.
The continuity equation for this charge reads
\begin{equation} \label{eq124}
\partial_t \rho_0  + \partial_{\bf x} \cdot {\bf j}_0({\bf x},t ) = e \delta({\bf x}-{\bf x}_0 ) \delta(t-t_0)
,\end{equation}
which is -- up to the factor $e$ -- the equation for the diffuson, since ${\bf j} = - D \partial_{\bf x} \rho $.
The charge density $\rho_0$ induces an electric potential, and an induced charge density,
which are related by the equation
\begin{eqnarray}
e \phi_{\rm ind}({\bf x}, t) & = & \int  \D {\bf x}_1 \D t_1 V_{\rm bare}({\bf x}, t; {\bf x}_1, t_1)
        [ \rho_0({\bf x}_1, t_1) + \rho_{\rm ind}({\bf x}_1, t_1) ] \\
   & =& \int \D {\bf x}_1 \D t_1 V({\bf x}, t; {\bf x}_1, t_1) \rho_0({\bf x}_1, t_1)
,\end{eqnarray}
where $V_{\rm bare}$ and $V$ are the bare and the screened interaction, and $\rho_{\rm ind }$ is the induced charge
density.
The convolution of the screened Coulomb interaction with a diffuson, as it appears in 
Eqs.~(\ref{eqCorr46}) and (\ref{eqCorr47}), is thus directly related to this induced electrical potential.
In some cases the latter is very conveniently determined from the continuity 
equation for the total charge density $\rho = \rho_0 + \rho_{\rm ind}$,
\begin{equation}
\partial_t \rho + \partial_{\bf x} \cdot {\bf j}({\bf x}, t) = e \delta({\bf x}-{\bf x}')\delta(t-t')
.\end{equation}
In the case of good screening this equation reduces to
\begin{equation}
\partial_{\bf x} \cdot {\bf j} = -\sigma \Delta \phi_{\rm ind}({\bf x},t) 
= e \delta({\bf x}- {\bf x}') \delta(t-t')
,\end{equation}
where $\sigma = 2 e^2 D N_0$ is the Drude conductivity.
The induced potential as a function of frequency and momentum then is
$e \phi_{\rm ind}({\bf q}, \omega ) = 1/(2 D N_0 q^2)$, as found in Eq.\ (\ref{eqCorr47})
 following different
arguments.

%
\subsection{Non-linear conductivity in films} 
\label{SecNonLinFilms}
\begin{figure}
\noindent
\begin{center}
\includegraphics[height=6.5cm]{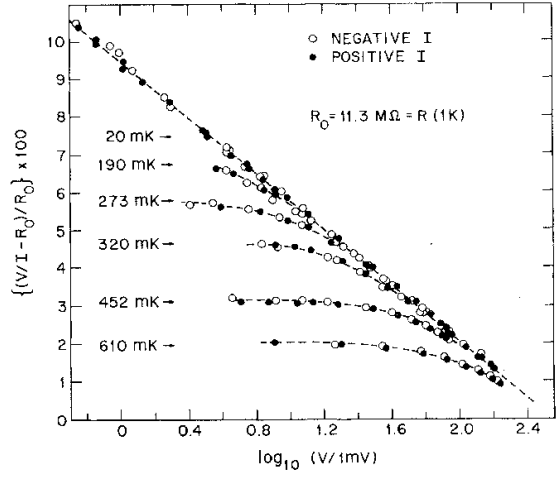}
\end{center}
\caption{\label{figCoul4}Resistivity as a function of voltage for a thin AuPd film, taken from \cite{dolan79}.
The resistance $R_0 = 11.3$ M$\Omega$ corresponds to $R_{\Box} \approx 4500$ $\Omega$.}
\end{figure}

In 1979 Dolan and Osheroff~\cite{dolan79} observed a logarithmic variation of the
resistivity of thin metallic films as a function of the applied voltage;
the experimental data are shown in Fig.~\ref{figCoul4}.
In order to  explain the experiment
Anderson {\it et al.}~\cite{anderson79} argued that the logarithm as a function of voltage is directly related to the
logarithm as a function of temperature (from weak localization in two dimensions)
since the dissipated power heats the electron gas. 
In the case of a strong electric field, the electron temperature is of the order of the
voltage drop on the relevant inelastic scattering length, the electron-phonon length,
{\it i.e}.\ $T \sim  e E L_{\rm e-ph}$, with 
$L_{\rm e-ph}= \sqrt{D \tau_{\rm e-ph}}$.
Assuming $L_{\rm e-ph}$ to be proportional to a power of the temperature, $L_{\rm e-ph} \sim T^{-p}$, electric
field and temperature are related as $T^{1+p} \sim E$ so that a logarithmic temperature dependence
of the linear resistivity causes a logarithmic voltage dependence.

Shortly after the first experiments, it was discussed whether heating is the only origin 
of the non-linear conductivity, or if
an electric field -- in analogy to a magnetic field --
can directly destroy weak localization via dephasing
\cite{tzuzuki81,kaveh81}.
The correct answer to the second question is ``no''
\cite{altshuler79a,bergmann82}, 
as one can easily verify by calculating the phase shifts of a pair of
time reversed paths.
In the presence of a vector potential an electron which propagates along a path ${\bf x}_1(t)$ or ${\bf x}_2(t)$
accumulates an additional phase
\begin{equation}
\varphi_{1,2} ={ e \over  \hbar} \int_0^\eta \D t' \dot{\bf x}_{1,2} \cdot {\bf A}.
\end{equation}
For a pair of closed time-reversed paths, ${\bf x}_2(t) = {\bf x}_1(\eta -t )$, and a static electric field,
${\bf A }(t) = -{\bf E} t$, the difference of the two phases vanishes, {\em i.e.}~there is 
no ``dephasing'' from a static electric field.

However, in a successive study of the current-voltage characteristic of gold films, 
Bergmann {\it et al.}~\cite{bergmann90} noted that the experimental data 
are not completely compatible with a pure heating model.
As a possible explanation of the experimental findings 
they suggested that the Coulomb interaction correction
to the resistivity leads to a non-ohmic contribution, namely 
\begin{equation} \label{eqCoulApp15}
\delta R_{\rm EEI}/R^2 =- { e^2 \over 4 \pi^2 \hbar } \ln\left[(k_B T)^2 + \alpha  \hbar D(eE)^2 \right/k_B T ] 
,\end{equation}
where $\alpha$ is a factor of the order one.

Indeed in formulating the phase shift argument for the interaction contribution,
a sensitivity to a static electric field cannot be excluded:
The interaction correction to the conductivity is related to the propagation of a
particle and a hole along closed paths.
In the absence of a vector potential the phases of particle and hole cancel, whereas 
in the presence of a vector potential the accumulated phase difference
is
\begin{equation} \delta \varphi =
  {e \over  \hbar} \int_{t_1-\eta}^{t_1} \D t' \dot{\bf x}_1 \cdot {\bf A }
- {e \over  \hbar} \int_{0}^\eta  \D  t'     \dot{\bf x}_2 \cdot {\bf A }
.\end{equation}
As is shown in Fig.~\ref{figCorr2} the relevant paths obey the relations
${\bf x}_1(t)={\bf x}_2(t)$ for
$0<t < t_1$ and ${\bf x}_1(t-\eta)={\bf x}_2(t)$ for $t_1 <t<\eta$,
which allows the phase difference to be written as
\begin{equation}
\delta \varphi =
{e \over \hbar} \int_{t_1-\eta}^0 d t' \dot{\bf x}_1 \cdot[ {\bf A}(t')- {\bf A}(t'+\eta) ]
.\end{equation}
For the particular case of a static electric field described by ${\bf A} = - {\bf E} t$, this
phase shift becomes
$\delta \varphi = {e \over \hbar} \eta ({\bf x}_2 - {\bf x}_1 ) \cdot {\bf E}$.
This suggests that the interaction correction should be sensitive to a static
electric field, leading  to a non-linear conductivity.
The quantitative calculation in fact results in \cite{raimondi99}
\begin{equation} \label{eqCoul33}
\delta R_{\rm EEI}/R^2 \approx  -{e^2 \over 2 \pi^2 \hbar }
\left( \ln k_B T + 1.62 {\hbar D (e E)^2 \over  (\pi k_B T )^3} \right)
,\end{equation}
verifying the non-ohmic behavior of the resistivity with the characteristic electric field scale
as suggested in \cite{bergmann90}. 
In order to derive Eq.~(\ref{eqCoul33}), a thermal distribution function with electron temperature $T$ has been assumed.
Following the steps in our calculation of the linear conductivity in section \ref{SecLin},
Eq.\ (\ref{eqCoul33}) is obtained 
from (\ref{eqCorr48}) in two dimensions under the condition $\hbar D (e E)^2 \ll  (\pi k_B T )^3$.
Note that the functional form (\ref{eqCoulApp15}) of the non-ohmic resistivity for strong electric fields 
cannot be confirmed theoretically.
We remark, finally, that further theoretical results have been given in
\cite{raimondi99,leadbeater00}, related to
strong electric fields, time dependent fields, one to three dimensions, and magnetic field effects.
%
\subsection{Gauge invariance} 
In the previous section we demonstrated that the non-linear electric field effect in the 
resistivity can be understood in terms of dephasing by calculating the phase shifts of the
relevant classical paths in the presence of a time dependent vector potential.
One may argue against this interpretation by noting that a vector potential 
can be gauged away
in such a way that the static electric field is described by a static scalar potential,
${\bf E}({\bf x}) = - \partial_{\bf x}\phi({\bf x})$.
A static scalar potential does no longer affect the diffuson, so that the argument of the phase difference along
two classical paths can no longer be used.
In order to clarify the situation,
we now verify explicitly the gauge invariance of the expression for the current density.
First one notices that  $\delta {\bf j}_{\rm Born} = -D \partial_{\bf x} \delta \rho$ is obviously gauge invariant.
For $\delta {\bf j}_V$, on the other hand, an explicit check is necessary.
Given the gauge transformation
\begin{eqnarray}
\label{eqCorr51}{\bf A} & \to & {\bf A} +\partial_{\bf x} \chi \\
\label{eqCorr52}\phi &\to & \phi - \partial_t \chi
\end{eqnarray}
the distribution function
and the diffuson
transform according to
\begin{eqnarray}
\label{eqCorr53} F(\eta; {\bf x}, t )  & \to &
F(\eta; {\bf x}, t )
\exp\{ -\I e[ \chi({\bf x}, t+ {\eta \over 2})- \chi({\bf x}, t-{ \eta\over 2} )]  \} \\
D^{\eta}({\bf x},t; {\bf x}', t' ) & \to &
D^{\eta}({\bf x},t; {\bf x}', t' )
\nonumber \\
&& \times \exp\{ -\I e [ \chi({\bf x}, t+{\eta\over 2} ) - \chi({\bf x}, t-{\eta \over 2})] \}\nonumber \\
\label{eqCorr54}
&& \times \exp\{ \I e [\chi({\bf x}', t'+{\eta\over 2})- \chi({\bf x}', t'-{\eta \over 2})]\}
. \end{eqnarray}
By applying the above transformation to $\delta {\bf j}_V$ as given in Eq.~(\ref{eqCorr46}), one can easily verify
that the function $\chi({\bf x}, t)$ drops out,
so that  the expression is manifestly gauge invariant.

For the specific example of a static electric field with
${\bf A} = - {\bf E} t$ we choose
$\chi({\bf x}, t) = t {\bf E} \cdot {\bf x} $. After the gauge transformation the electric field
appears in the distribution function,
$\tanh(\epsilon/ 2 T)  \to \tanh [ (\epsilon - \mu_{\bf x}) / 2 T ]$, $ \mu_{\bf x} = e {\bf E} \cdot {\bf x}$,
but not in the diffuson.
We conclude that although the contribution of the Coulomb interaction to the
current density is gauge invariant, the interpretation of the non-linear electric field effects
depends on the actual choice of the potentials ${\bf A}$ and ${\bf \phi}$.
With ${\bf E}= - \partial_t {\bf A}$, $ \phi =0$ we would interpret the non-linear
conductivity as due to a phase effect.
With ${\bf E} = -\partial_{\bf x} \phi$, ${\bf A}=0$ the origin of the
non-linear conductivity can be attributed to the varying local chemical potential
which is felt by the traveling particle, and the virtual
hole created and absorbed in the intermediate states.
%
\subsection{Electron dephasing} 
The dephasing time sets the scale over which an electron propagates without loosing phase coherence, and
determines the amplitude of the quantum interference corrections to the conductivity.
Since elastic scattering does not destroy the electronic phase coherence, inelastic scattering
must be hold responsible.
Dephasing has been studied in much detail for weak localization
\cite{altshuler82,fukuyama83,eiler84,stern90} (for a recent experimental review see \cite{lin02}),
and also in the context of universal conductance fluctuations \cite{hoadley99,aleiner02b}.
The amplitude of the interaction correction to the conductivity, on the other hand, is set
by the thermal time $\tau_T = \hbar/k_B T$.
In most cases the dephasing time is much longer than the thermal time.
In some experiments, however, an extraordinarily strong phase breaking has been reported:
Recently a low temperature saturation of $\tau_\varphi$ in gold wires \cite{mohanty97,mohanty98} has attracted much
attention
\cite{golubev98,altshuler98,aleiner99,cohen99,zawadowski99,imry99,gougam00,kirkpatrick02}.
Furthermore, in a number of cuprates,
the dephasing rate decreases only slowly with decreasing temperature
\cite{jing91,hagen92,fournier00}.
For example in
$\rm Bi_2Sr_2CuO_6$, \cite{jing91}, the dephasing rate varies as $1/\tau_\varphi \sim T^{1/3}$,
with $\tau_\varphi$ much shorter than $\tau_T$.
In such a case phase breaking may become relevant also in the interaction contribution to the conductivity.

Unfortunately, in the experiments cited above, it is not clear which microscopic mechanism is responsible for
the strong phase breaking.
Nevertheless we believe it is important to answer the following questions:
(i) Is phase breaking relevant in the interaction contribution to the conductivity?
(ii) If yes, is the phase breaking rate which is relevant in the interaction contribution to the
conductivity the same as for weak localization? While
Castellani {\it et al.}~\cite{castellani86} came to the conclusion that the answer is ``yes'' to both
questions,
Raimondi {\it et al.}~\cite{raimondi99} who reexamined the problem,
confirmed the suggestion that
dephasing is relevant also in the interaction contribution to the conductivity, but
found a different dephasing time for the latter compared to weak localization.
After some historical remarks we will
give a semi-quantitative summary of the analysis carried out in \cite{raimondi99}.

While it is clear that inelastic scattering contributes to dephasing, the precise way this happens
is less obvious.
It was first noted by Schmid \cite{schmid74} that the inelastic quasi-particle scattering time is enhanced
in the presence of disorder. The dephasing time
was initially assumed to be identical to the inelastic quasi-particle scattering
time \cite{abrahams81},
thus the inverse dephasing time was assumed to be
$\propto T \ln T$ in two dimensions.
Some time later, Fukuyama and Abrahams \cite{fukuyama83}
reexamined the problem in terms of standard diagrams and calculated the ``mass''
that develops in the particle-particle propagator (the cooperon) in the
presence of Coulomb interaction. They also found an inverse time proportional
to $T \ln T$. Furthermore a mass term
also appears in the particle-hole propagator (the diffuson) \cite{castellani86}.
The mass in the particle-hole propagator turned out to coincide with that in
the particle-particle propagator.
However, calculating a mass is not sufficient in order to draw final
conclusions;
Altshuler, Aronov, and Khmelnitskii \cite{altshuler82}, for example,
calculated the dephasing time via a path-integral approach
and predicted an inverse dephasing time proportional to $T$,
in contrast to \cite{abrahams81}.
This result -- see also \cite{eiler84} -- has been confirmed in many experiments. The approach of
Altshuler, Aronov, and Khmelnitskii has been extended later in order to determine
the relevant dephasings times for the universal conductance fluctuations \cite{aleiner02b},
and in the particle-hole channel \cite{raimondi99}.

The general idea is that the
phase of an electron which is propagating along a path ${\bf x}_{1,2}(t)$
is shifted by $ \varphi_{1,2} = {e } \int \D t \, \dot {\bf x}_{1,2}(t') \cdot  {\bf A}(t')$.
In the presence of fluctuating internal fields this results in phase fluctuations
which finally destroys phase coherence.
In thermal equilibrium the internal electric field fluctuations
are given by
\begin{eqnarray}
e^2 \langle E^i E^j  \rangle_{\bf q, \omega} & = &  q^i q^j{\I \over 2} V^K({\bf q}, \omega) \\
& \approx &  - q^i q^j  {2T \over \omega } \Im V^R({\bf q}, \omega)
,\end{eqnarray} 
where $V^K({\bf q}, \omega)$, $V^R({\bf q}, \omega)$ are the Keldysh and the
retarded Coulomb interaction, and low frequencies
($\omega \ll T$) have been assumed. 
The assumption of low frequencies is essential here since only in this frequency regime the
classical field fluctuations, {\em i.e}.\ the Keldysh component of the interaction, dominates whereas 
at higher frequencies the nontrivial structure of the interaction in Keldysh space has to be taken into account.
It has been shown by Fukuyama and Abrahams \cite{fukuyama83} and  Castellani {\it et al.}\ \cite{castellani84}
that the high frequency contributions of the Coulomb interaction renormalize parameters 
like the diffusion constant
but do not contribute to dephasing. It is therefore reasonable to neglect these high frequency contributions in the following.
The particle-particle propagator (cooperon) and the 
particle-hole propagator (diffuson) in the presence of the fluctuating field can then be written as
a path-integral: 
\begin{eqnarray}
C^t_{\eta, \eta'}({\bf x}, {\bf x}') & = & {1 \over 2} \int_{{\bf x}'_{\eta'} = {\bf x}' }^{{\bf x}_\eta= {\bf x}}
{\cal D} {\bf x}_{t_1} \exp-( S_0^{\rm p.p.}+ S_1^{\rm p.p.}) \\[0.1cm]
D^\eta_{t, t'}({\bf x}, {\bf x}') & = & \int_{{\bf x}'_{t'} = {\bf x}' }^{{\bf x}_t= {\bf x}}
{\cal D} {\bf x}_{t_1} \exp-( S_0^{\rm p.h.}+ S_1^{\rm p.h.} )
\end{eqnarray}
with
\begin{eqnarray}
&&S_0^{\rm p.p.} = \int_{\eta'}^{\eta} \D t_1 { \dot {\bf x}_{t_1}^2 \over 2 D}, \quad
S_1^{\rm p.p.} = -\I e \int_{\eta'}^\eta \D t_1 \dot{\bf x}_{t_1} \cdot [{\bf A}(t+t_1/2)+ {\bf A}(t-t_1/2) ]  \\
&& S_0^{\rm p.h.} = \int_{\eta'}^{\eta} \D t_1 { \dot {\bf x}_{t_1}^2 \over 4 D}, \quad
S_1^{\rm p.h.} = -\I e \int_{\eta'}^\eta \D t_1 \dot{\bf x}_{t_1} \cdot [{\bf A}(t_1 + \eta/2 )- {\bf A}(t_1 - \eta/2) ]
.\end{eqnarray}
In both cases the $S_1$ term describes the additional phase due to the vector potential ${\bf A} $.
Since we have to consider the phase difference of pairs of classical paths, 
the vector potential appears twice in  $S_1$.
In the case of the cooperon the two paths are the time reversed of the other,
{\em i.e}.,  to a velocity $\dot {\bf x}$ on path
``$1$'' corresponds a velocity $-\dot {\bf x}$ on path ``$2$'' and in the expression for the phase difference 
the sum of two vector potentials appears.
In the case of the diffuson the two paths are traversed in the same direction, but at different
times. Therefore the  minus sign between the two vector potentials remains, and
in particular when the two paths are traversed at the same time ($\eta =0 $) the phase difference 
vanishes.
By averaging the phase factor $\exp(- S_1 )$ over the electric field fluctuations, using
the relation
\begin{equation}
\langle \exp ( \I \delta  \varphi  ) \rangle = \exp \left(  - {1\over 2} \langle (\delta  \varphi )^2\rangle \right)
\equiv \exp (-S)
,\end{equation}
one finds
\begin{eqnarray}
S^{\rm p.p.} & = & {1 \over 2} \int_{-\eta}^\eta \! \D  t_1 \int_{-\eta}^\eta \! \D t_2 
\int_{-T}^T {\D \omega \over 2 \pi} \sum_{\bf q}{ 2 T \over \omega }\left[ - \Im V^R({\bf q}, \omega) \right]  \cr
\label{eqApp24}
  &\times & \exp\left\{ \I {\bf q}\cdot [{\bf x}_1(t_1)-{\bf x}_1(t_2)] \right\}
 \left[ \cos\left(\omega {t_1 - t_2  \over 2 } \right) - \cos \left(\omega {t_1 + t_2 \over 2}  \right) \right]
\end{eqnarray}
in the case of the cooperon, and
\begin{eqnarray}
S^{\rm p.h.} & = &  \int_{0}^\eta \! \D  t_1 \int_{0}^\eta \! \D t_2 
\int_{-T}^T {\D \omega \over 2 \pi} \sum_{\bf q}{ 2 T \over \omega }\left[ - \Im V^R({\bf q}, \omega) \right]  \cr
\label{eqApp25}
  &\times & \exp\left\{ \I {\bf q}\cdot [{\bf x}_1(t_1)-{\bf x}_1(t_2)]\right\} \E^{-\I \omega (t_1 - t_2) }
 \left[ 1-\cos( \omega \eta ) \right]
\end{eqnarray}
in the case of the diffuson; in both $S^{\rm p.p.}$ and $S^{\rm p.h.}$ the
velocities $\dot{\bf x}(t_{1,2})$ were removed by a partial integration, 
and the relation between the two relevant paths ${\bf x}_1(t)$ and ${\bf x}_2(t)$ was exploited.
The expressions for $S^{\rm p.p.}$ and $S^{\rm p.h.}$ appear to be similar, except for the time dependent factors
in the second line; 
this leads to different dephasing times in the particle-particle and particle-hole channels. 
In two dimensions, for example, the final results are
\begin{equation}
S^{\rm p.p.} \approx  (T \eta ) { R_\Box \over h/e^2 } \ln (T \eta ) ; \, \, \, \,
{1 \over \tau_\varphi} \approx T {R_\Box \over h/e^2 } \ln\left(  {h/e^2 \over R_\Box} \right) 
\end{equation}
and
\begin{equation}
S^{\rm p.h.} \approx  (T \eta)   { R_\Box \over h/e^2 } \ln ( D \kappa^2 T \eta^2 ); \, \, \, \,
{1 \over \tau_\varphi} \approx T { R_\Box \over h/e^2 } \ln \left[ { D \kappa^2 \over T } 
 \left( {h/e^2 \over R_\Box}  \right)^2 \right]
,\end{equation}
where $\kappa$ is the inverse screening length, and $R_\Box$ is the sheet resistance.
The dephasing time is here determined from the condition
$S(\eta \!  =   \! \tau_\varphi)=1$. In the particle-particle channel the standard result 
of Altshuler, Aronov and Khmelnitskii \cite{altshuler82,eiler84,stern90} is confirmed. 
The dephasing time in the particle-hole channel is finite,
different from the one in the particle-particle channel, and 
is identical to the inelastic scattering rate in the two-particle propagators (diffuson or cooperon), as first
calculated by Fukuyama and Abrahams \cite{fukuyama83,blanter96}.

The low frequency electric field fluctuations considered here, cannot explain the strong
phase breaking observed in \cite{mohanty97,jing91}. However we expect similar results 
for other phase breaking mechanisms: 
We suggest that a mechanism leading to strong dephasing in the particle-particle channel will also 
cause strong dephasing in the particle-hole channel. 
In particular when $\tau_\varphi$ becomes comparable to or shorter than the thermal time $\tau_T$,
the relevant time scale which sets the amplitude of the Coulomb interaction contribution to the conductivity
will be $\tau_\varphi$ instead of $\tau_T$. 
This is consistent with the experiments:
In the gold wires of \cite{mohanty97} the dephasing rate saturated below $T\sim 1$ K to values of the
order of $\hbar/\tau_{\varphi} \sim 1$-$10$ mK. In the samples with the strongest phase breaking a saturation
of the interaction correction to the conductivity has been observed below $T\sim 100$ mK \cite{mohanty98}.
In $\rm Bi_2 Sr_2 Cu O_6$, a compound with a single $\rm Cu O_2$ plane per unit cell, the in-plane zero field resistivity
increases as $\ln T$ below $\sim$18 K, consistent with quantum interference effects in two dimensions.
The shape of the orbital magneto-resistance is well fitted
by the weak localization expression \cite{jing91},
but with an unexpected large dephasing rate which varies as $T^{1/3}$.
The spin component of the magneto-resistance varies at low fields as
$ [R(B)- R(0)]/R(0) \sim ( B/B_s )^2$.
From the standard theory for the Coulomb interaction in the spin triplet channel \cite{altshuler80b,lee85} 
one would expect such a quadratic magnetic field dependence with a magnetic field scale which is
linear in the temperature, $B_s \sim 1/\tau_T $. Experimentally
the magnetic field scale varies as $T^{0.4}$, which is 
close to the temperature variation of the dephasing rate. This 
suggests that the origin of the 
spin component of the magneto-resistance might indeed be the Coulomb interaction contribution to the conductance,
but in the presence of an up to now unidentified phase breaking mechanism.
%
%
%

\section{The Coulomb interaction in nanostructures} 
\label{SecNanNo}
Quasiclassical methods -- as pioneered by Nazarov -- 
allow description of interaction  effects in a unified way.
For instance, in the following subsection we will show that Coulomb blockade physics in 
small tunnel junctions and disorder-induced zero-bias anomalies in diffusive conductors
are closely related. 
More generally the question arises how the Coulomb interaction affects the transport in
small systems with diffusive electron motion, where interfaces provide an 
additional source of scattering. Typical situations will be examined in detail 
in the subsequent subsections.
\subsection{Coulomb interaction and interface conductance}
\label{SecCoulInterface}
The Coulomb interaction plays an important role
not only in diffusive transport, but also for transport 
across interfaces.
For example, the Coulomb interaction reduces the density of states near the Fermi energy,
\cite{altshuler79,altshuler80a,altshuler84,finkelstein83,castellani84,levitov97,kopietz98,kamenev99}
which leads to a suppression of the transmission of tunnel junctions, 
and to a non-linear current-voltage characteristic at low bias.
On the other hand Coulomb blocking of tunneling may also be due to interface charging effects
\cite{ambegaokar82,ben83,ho83,schon90,ingold92}.
These two classes of phenomena were described in a unified way for the first time by Nazarov
\cite{nazarov89a,nazarov89b}, and later by various authors 
\cite{levitov97,kamenev96,sukhorukov97,minkov99,pierre01,beloborodov01,rollbuhler01}.

In this subsection we will show how to construct the theory within the quasiclassical scheme.
Whereas in the above cited references it is assumed that the
electronic system on both sides of the interface is in thermal equilibrium, there
is no such assumption in the approach we present here.
The idea is to start with the boundary conditions for the quasiclassical Green's function,
derive an expression for the current through the interface, and at the end average over the fluctuations
of the internal electric field.

Zaitsev's boundary condition \cite{zaitsev84,shelankov84},
which generalizes the matching condition (\ref{eq55a}) to situations where 
$g^{R,A} \ne \pm 1$, reads
\begin{equation} \label{eqZaitsev}
 \check a \left[ R - R \check a \check a  + {T \over 4}(\check s({\bf x}_-)- \check s ({\bf x}_+ ))^2 \right] =
{T \over 4} [ \check s({\bf x}_+) , \check s({\bf x}_-) ]
\end{equation}
with
\begin{eqnarray}
 \check  s & = & {1\over 2} \left[ \check g ( \hat {\bf p}) + \check g (\tilde  {\hat{\bf p}})  \right] \\
 \check a & = &  {1\over 2}  \left[ \check g ( \hat {\bf p}) - \check g (\tilde  {\hat{\bf p}})  \right]
.\end{eqnarray}
The antisymmetric combination of Green's functions $\check a$ is continuous across the interface.
In the absence of electronic interactions ($g^{ R, A} = \pm 1 $), we recover Eq.~(\ref{eq55a}).
In the interacting case, where $g^{R,A} \ne \pm 1$, we restrict our considerations to the limit
of low interface transparency, $T/R \ll 1 $ . 
Then the structure of the boundary conditions becomes simple, 
 \begin{equation}
a^K  = {T \over 4} [\check s({\bf x}_+),  \check s({\bf x}_-)]^K
,\end{equation} 
and the current density across an interface which connects two dirty pieces of metal is given by
\begin{eqnarray} \label{eq112}
\hat {\bf n} \cdot {\bf j} & =  &
 e \pi N_0 \Gamma {1\over 2} \int {\D \epsilon \over 2 \pi }
\langle \left[ \check g_s({\bf x}_+), \check g_s({\bf x}_-) \right]^K  \rangle_{\phi^{\rm int}} \\
\Gamma &= & v_F {1 \over 2}  \int_0^1 \D (\cos \theta) \cos \theta  \, 
 T( \cos \theta )
.\end{eqnarray}
In the absence of interactions the tunneling conductivity of the interface is
given by $\sigma_T = 2 e^2 N_0 \Gamma$.
The remaining technical difficulty in order to evaluate the current 
through an interface in the presence of interactions 
is the average over the internal field fluctuations.
Solutions are known only in some special situations;
one of these is when the two sides of the interface are uncorrelated 
such that the average over the internal field
fluctuations factorizes. The current can then be written in a familiar form 
in terms of the local density of states 
$N(\epsilon, {\bf x})$ as
\begin{eqnarray}
\hat {\bf n} \cdot {\bf j} & =  &
 e {   \Gamma \over N_0} \int {\D \epsilon  }
\left[ F(\epsilon, {\bf x}_- )- F(\epsilon, {\bf x}_+) \right]
N(\epsilon, {\bf x}_+) N(\epsilon, {\bf x}_- ) \\
N(\epsilon, {\bf x}) & = & \pm N_0 \Re \langle  g^{R,A}(\epsilon, {\bf x}) \rangle_{\phi^{\rm int}}
.\end{eqnarray}
In general, however, such correlations do exist due to the interaction between the electrons
on opposite sides of the interface.

Another situation where Eq.~(\ref{eq112}) 
can be evaluated explicitly
is when the gradients in the kinetic equation on both sides of the interface are negligible,
{\em i.e}.\ when the diffusive motion of the charge carriers is negligible.
In this limit standard Coulomb blockade theory ($P(E)$ theory, \cite{ingold92a}) is recovered as we demonstrate here.
The question under which circumstances the
gradient terms in the kinetic equation can be neglected will not be addressed here: For a discussion of this issue see
the appendix of \cite{ingold92a}, and also subsection \ref{SecDiffDot}.
Neglecting the gradient term, the kinetic equation in the presence of the Coulomb interaction
can be solved explicitly, and the Green's function reads
\begin{equation}
\check g_{t_1 t_2} ({\bf x} ) =  \E^{\I \check  \varphi (t_1, {\bf x}) } 
\check g_{0,t_1 t_2}({\bf x})
 \E^{-\I \check  \varphi (t_2, {\bf x}) } 
,\end{equation}
with $ \partial_t \check \varphi = e \check \phi_{\rm int} $, and $\check g_0$ is the Green's function
in the absence of the interaction.
Since we assume Gaussian fluctuations for the internal field $ \phi_{\rm int}$, it is clear that 
the average over $\phi_{\rm int}$ can be carried out for both the Green's functions
and 
the current density, Eq.~(\ref{eq112}).
The final result for the current density can be written in the form 
\begin{eqnarray} \label{eq151a}
\hat {\bf n} \cdot {\bf j} & =  &
e 2 \Gamma N_0 \int {\D \epsilon  }  \int \D \omega  P( \omega ) \cr
&& \times 
\left\{ f(\epsilon, {\bf x}_+ )  [ 1 - f( \epsilon - \omega, {\bf x}_- ) ]  
-f(\epsilon, {\bf x}_- ) [ 1 - f( \epsilon - \omega,  {\bf x}_+) ]  \right\}
\end{eqnarray}
with 
$F(\epsilon, {\bf x}) = 1 -2 f(\epsilon, {\bf x})$.
The factors $f(\epsilon, {\bf x}_+ )  [ 1 - f( \epsilon - \omega, {\bf x}_- ) ]$
describe the tunneling of a charge from 
an occupied state on the ``$+$''-side of the junction to an unoccupied state on the ``$-$''-side.
In the absence of interactions the function $P(\omega)$ is a delta function centered at zero, and the
usual expression for the tunnel current is obtained.
In the presence of interactions a tunneling electron may exchange energy with the environment,
and $P(\omega)$ may be interpreted as the probability to transfer the energy $\omega$ to the
environment \cite{ingold92a}.
The function $P(\omega ) $ is given by
\begin{eqnarray}
P( \omega ) & = & {1\over 2 \pi } \int \D t \E^{\I \omega t} \E^{J(t)} \\
{J(t)}   & = &   
\langle (\varphi_1(t)-\varphi_1(0))\varphi_1(0) \rangle_{\phi_{\rm int}} 
+ \langle  \varphi_1(t)\varphi_2(0) \rangle_{\phi_{\rm int}} 
- \langle  \varphi_2(t)\varphi_1(0) \rangle_{\phi_{\rm int}} 
,\end{eqnarray} 
where $\varphi_{1,2}$ is the phase difference between the left and right leads,
 $\varphi_{1,2}= \varphi_{1,2}({\bf x}_-) - \varphi_{1,2}({\bf x}_+)$;
the index $1,2$ refers to the Keldysh space.
In thermal equilibrium and expressing the phase/voltage fluctuations over the junction in terms of
an impedance $Z$ one has
\begin{equation}
\int \D t  \E^{\I \omega t}  \langle \varphi_1(t)\varphi_2(0) \rangle_{\phi_{\rm int}} = {1\over 2}
 { e^2 Z \over  \omega}
;\end{equation}
the function $J(t)$ is determined as
\begin{equation} \label{eq155a}
J(t) =  \int {\D \omega \over \omega} {\Re Z \over 2\pi/ e^2  } 
\left\{ 
\left[\cos(\omega t ) - 1   \right] \coth\left( \omega \over 2T \right) - \I \sin(\omega t )
\right\}
.\end{equation}
As an example, to mimic the electrodynamic environment of a tunnel junction, assume the impedance $Z$ to be
$Z=1/( - \I \omega C + R^{-1})$, {\em i.e}.\ a capacitance in parallel with a resistor.
For a high impedance, $R \gg h/e^2$, when the charge relaxation process becomes slow, one obtains 
\begin{equation}
J(t) = - \I E_c t - E_c T t^2
,\end{equation}
where $E_c=e^2 /2 C$ is the charging energy.
In the zero temperature limit $P(\omega)$ reduces to a delta function,
\begin{equation}
P(\omega) = \delta( \omega - E_c )
,\end{equation}
such that a minimum voltage difference $E_c/e $ is needed
for an electron to tunnel
through the junction.
The current-voltage characteristic is given by
\begin{equation}
I(V) = G_T( V - E_c/e ) \Theta( eV - E_c ), \quad G_T = 2 e^2 N_0 \Gamma {\cal A}.
\end{equation}
For more applications of the standard Coulomb blockade theory we refer to the literature, {\em e.g}.\
\cite{ingold92}.

We close this subsection by highlighting how the zero-bias Altshuler-Aronov anomalies arise.
These are present in situations
where the gradient in the kinetic equation cannot be neglected.
Since under this condition 
an exact expression for the Green's function near the junction is not
known,
we restrict ourselves to a pertubative expansion in the internal field $\check \phi_{\rm int}$,
with the result
\begin{eqnarray} \label{eq163a}
\hat {\bf n} \cdot \delta {\bf j} & =  &
- e \pi N_0 \Gamma {1\over 2} \int {\D \epsilon \over 2 \pi }
\left[ F(\epsilon, {\bf x}_-) - F( \epsilon,  {\bf x}_+ ) \right] \nonumber  \\[0.1cm]
 &&\times \Big[
\Re   \langle \delta g^K({\bf x}_+ ) [\delta g^Z({\bf x}_+)- \delta g^Z({\bf x}_-) ] 
     \rangle_{\phi^{\rm int}}  \nonumber \\[0.1cm]
&&  +  \Re \langle \delta g^K({\bf x}_- ) [\delta g^Z({\bf x}_-)- \delta g^Z({\bf x}_+) ] 
      \rangle_{\phi^{\rm int}} \Big]
,\end{eqnarray}
where $\delta g^K$ and $\delta g^Z$ are the correction to the Green's functions
due to the fluctuating field.
The relevant correlation function has been given in Eq.~(\ref{eq106}),
and we find
\begin{eqnarray}
\hat {\bf n} \cdot \delta {\bf j} & = &
{1 \over 2 e} \sigma_T  \int \D \epsilon \left[ F(\epsilon, {\bf x}_-) - F( \epsilon,  {\bf x}_+ ) \right]
     \Im \int {\D \omega \over 2 \pi} \int \D {\bf x}_1 
   F(\epsilon - \omega, {\bf x}_1) \nonumber  \\[0.1cm]
&& \times\left\{  \rho_0({\bf x}_+, \omega; {\bf x}_1 )
   \left[ \phi_{\rm ind}({\bf x}_1, \omega; {\bf x}_+) - \phi_{\rm ind}({\bf x}_1, \omega;{\bf x}_-)\right] 
    \right. \nonumber  \\[0.1cm]
\label{eq164a}
&& \left. + \rho_0({\bf x}_-, \omega; {\bf x}_1 )
   \left[ \phi_{\rm ind}({\bf x}_1, \omega; {\bf x}_-) - \phi_{\rm ind}({\bf x}_1, \omega;{\bf x}_+)\right]
    \right\}
,\end{eqnarray}
where $\sigma_T$ is the tunneling conductivity in the absence of interactions,
$\rho_0({\bf x}, {\bf x}')$ and $\phi_{\rm ind}({\bf x}, {\bf x}')$ 
are the charge density and the induced electrical potential at ${\bf x}$ due to a charge which is generated at ${\bf x}'$,
compare subsection \ref{SecInducedPotential}.
When interactions are taken into account only on one side of the interface,
the correction to the current is controlled by the density of states.
For example in three dimensions the correction to the density of states is proportional to $\sqrt{\epsilon}$,
giving rise to a $\sqrt{V}$ of the tunneling conductivity as a function of voltage \cite{altshuler85}.
However, in the more general situation, as pointed out by Nazarov \cite{nazarov89a,nazarov89b}, correlations
in both leads and between the leads have to be taken into account.

For further experiments where the perturbative zero bias anomalies in the tunnel conductivity have been observed,
see for example \cite{sukhorukov97,minkov99,pierre01}.
We will work out the Coulomb correction (\ref{eq164a})
to the interface current for a specific geometry in section
\ref{SecShortWire}.

%

\subsection{Non-linear conductivity in wires} 
\label{SecLongWire}
In this and the following two subsections we will discuss the
contribution of the Coulomb interaction to the conductivity in small systems. 
We would like to mention that size effects in the quantum corrections to the {\em linear} conductivity
were studied in detail in the mid-eighties 
\cite{altshuler84,masden82,santhanam87}. 
In contrast, we will focus here on the case
of a large applied voltage.  We will
successively decrease the size of the systems under consideration.
This subsection is devoted to wires which may be shorter than the
inelastic scattering length, but which are still longer than the thermal diffusion length
$L_T = \sqrt{\hbar D /k_B T}$.
This is then followed by subsection \ref{SecShortWire} where we consider system 
sizes of the order of $L_T$. There it will be essential to consider the discreteness of the diffusive 
modes, and we will also allow for resistive interfaces between the wire and the leads.
Finally, in subsection \ref{SecDiffDot} we will consider the situation of a small, fully coherent
 piece of metal beeing connected to two large reservoirs.
\begin{figure}
\noindent
\hspace{1.5cm}{\includegraphics[width=12cm]{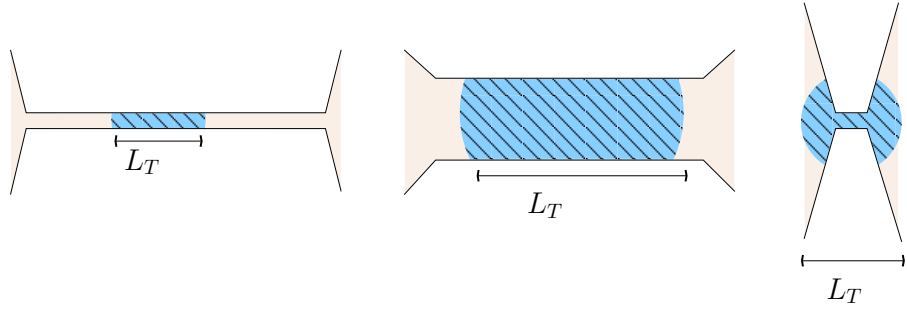}}
\caption{The geometries considered in subsections \ref{SecLongWire} -- \ref{SecDiffDot}. 
Section \ref{SecLongWire} is devoted
to wires which are much longer than the thermal diffusion length $L_T$. In Sec.~\ref{SecShortWire}
$L_T$ is of the order of the system size; both in Sec.~\ref{SecLongWire} and Sec.~\ref{SecShortWire} 
we assume ideal leads.
In section \ref{SecDiffDot} a diffusive point contact is considered, where the system size is much smaller
than $L_T$ such that  the phase coherent volume extends into the leads, which then have to be taken into account
explicitly.
}
\end{figure} 

We start with a long, thin wire,
for which the Coulomb correction to the resistivity, including the leading non-linear voltage dependence,
in analogy to Eq.~(\ref{eqCoul33}) for a thin film, is given by \cite{leadbeater00}
\begin{equation} \label{eqCoul34}
\delta R_{\rm EEI} /R^2 \approx {e^2 \over \pi^2 \hbar} {L_T \over L }
\left( 4.91 -0.21 {\hbar D (eV/L)^2 \over (k_B T)^3 } \right)
,\end{equation}
where $V= EL$ is the applied voltage.
Again, this result has been 
obtained under the assumption of a thermal distribution function with a constant temperature $T$.
Whereas this is reasonable for macroscopic samples, it fails, however,
in samples which are shorter than the electron-phonon scattering length,
in which case  
the Coulomb interaction correction to the conductivity far from equilibrium
\cite{nagaev94,gutman01,schwab01} has to be considered.

For an evaluation of the current-voltage characteristic, the diffuson and the distribution function in the
wire are required.
The diffuson is found by solving the differential equation (\ref{eqCoul22}) with the condition that
the derivative normal to an insulating boundary (i.b.) vanishes, {\em i.e.}
\begin{equation} \label{eqApp15}
( \hat {\bf n}\cdot \nabla_{\bf x} ) D({\bf x},{\bf x}'){\big |}_{{\bf x}\, \in {\rm \, i.b.} } =0,
\end{equation}
while the diffuson itself vanishes at a metallic boundary (m.b.),
\begin{equation} \label{eqApp16}
D( {\bf x},  {\bf x}'){\big |}_{{\bf x} \,\in {\rm \, m.b.} } =0
.\end{equation}
The first condition means that no current flows through the interface, the
second condition corresponds to the assumption that an electron 
arriving at the metallic boundary escapes into the leads with zero probability to come back into the wire.
Furthermore it is assumed that the left and right leads of the wire are in thermal equilibrium,
\begin{equation}
F(\epsilon, {\bf x}){\big |}_{{\bf x} \, \in  \,l.l.,r.l.} = \tanh\left({\epsilon \pm |e| V/2 \over 2T} 
\right).
\end{equation}
By solving the kinetic equation it is found that the distribution function
depends on the various relaxation mechanisms governing the collision
integral \cite{nagaev92,nagaev95,kozub95,naveh98,altshuler98},
and we distinguish three regimes:
\begin{itemize}
\item[a)]
When the system is much longer than the electron-phonon scattering length, 
$L \gg L_{\rm e-ph}$, the distribution function acquires the equilibrium form with a local chemical
potential and temperature, 
\begin{equation}\label{eqCoul40}
F(\epsilon, x) = \tanh \left( { \epsilon +  |e| V(L-2x)/2L \over 2 T_e(x,V) } \right)
,\end{equation}
where $x=0 \dots L$ is the distance from the left lead. 
The local electron temperature $T_e(x)$ may be determined from an energy balance argument, 
assuming that the dissipated power equals the gradient of the heat flow. 
In the limit considered here the heat flow is dominated by the phonons.
For a stationary temperature $T_e(V)$ the Joule heating power $P_{\rm in}= \sigma {\bf E}^2$
equals the power which is transferred into the phonon system, $P_{\rm out}$.
For weak heating one has
$P_{\rm out} = c_V \Delta T /\tau_{\rm e-ph}$, where $c_V$ is the electron specific heat, 
$\Delta T$ is the difference between the
 electron and phonon temperatures, and $\tau_{\rm e-ph}$ is the relevant energy relaxation rate.
For strong heating, on the other hand, the effective electron temperature is of the order of the
voltage drop over a phonon length,
$T_e \sim |e| V L_{\rm e-ph}/L$.
If the ``hot'' phonons escape ballistically into the substrate, 
the temperature in the bulk of the wire does not depend on the position $x$ but is voltage dependent, $T_e=T_e(V)$.
By neglecting the region near the leads, where the temperature rises from $T_{\rm lead}$ to $T_e(V)$,
Eq.~(\ref{eqCoul34}) is recovered for the voltage dependent resistivity.
\item[b)]
When $L_{\rm e-ph} \gg L \gg L_{\rm in}$ one still expects a distribution function near local equilibrium,
due to electron-electron scattering. 
The local temperature is determined from the relation
$ \sigma E^2 = $$- \nabla [ \kappa \nabla T_e(x) ] $,
where $\kappa$ is the thermal conductivity.
Using the Wiedemann-Franz law, 
$\kappa =({\pi^2/ 3}) $ $T_e \sigma ( k_B/e)^2 $, 
the temperature profile in the wire is determined as
\begin{equation}
T_e^2(x) = T_{\rm lead}^2  + {3 \over \pi^2}\left( {eV \over L } \right)^2 x(L-x)
.\end{equation}
\item[c)] 
In the absence of inelastic scattering,
realized when $L_{\rm in } \gg L$, the distribution function is a linear superposition of the 
distribution function of the leads,
\begin{equation} \label{eqCoul42}
F(\epsilon, x) =  \left[ (L-x) F(\epsilon, 0 ) + x F(\epsilon, L)\right]/L
.\end{equation}
\end{itemize}
The resistance as a function of voltage in regime a) 
is given in Eq.\ (\ref{eqCoul34}).
This equation applies when the voltage drop over a thermal diffusion length
is smaller than the temperature, $ e V L_T/L < T$.
Since we assume that $L_{\rm e-ph} > L$ the electron temperature as a function of voltage 
rises so fast that this condition always holds.
We compare now the 
relative importance of both heating and non-heating effects for the non-linear resistivity.
From the linear conductivity and the increase of temperature due to the applied 
voltage as discussed above,
we find at low voltage 
\begin{equation}
 \delta R_{\rm heating} / R^2 \approx - 4.91 {e^2 \over \pi^2} {L_T \over L}  
 { D ( e V/L)^2 \over T^2 (1/ \tau_{\rm e-ph} )}
,\end{equation} 
which has to be compared with the corresponding change of the resistance due to
non-heating effects:
\begin{equation}
 \delta R_{\rm non-heating} / R^2 \approx - 0.21{e^2 \over \pi^2} {L_T \over L}{ D ( e V/L)^2 \over T^3 }
.\end{equation}
One observes that the heating contribution to the non-linear resistivity is 
larger 
by a factor
$T \tau_{\rm e-ph}$.
Thus the non-heating effects are difficult to observe experimentally,
since usually $1/\tau_{\rm e-ph} \ll T$.

\begin{figure}
\noindent
\hspace{2cm}{\includegraphics[width=8cm]{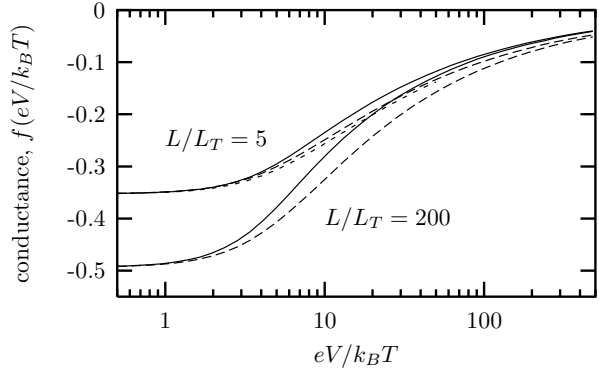}}
\caption{Interaction correction to the conductance $\delta I_{\rm EEI}/V$ for a mesoscopic wire
as a function of voltage, taken from \cite{schwab01}. $\delta I_{\rm EEI}/V$ is plotted in units
of $(e^2/\hbar) L_T/L$.
The full line corresponds to the non-equilibrium distribution function (\ref{eqCoul42}).
The line with long dashes corresponds to
the local equilibrium distribution function (\ref{eqCoul40}) with $x$-dependent temperature.
The line with short dashes ($L/L_T = 5$)
is the non-linear conductivity due to the heating contribution only, Eq.~(\ref{eqHeating}).}
\label{figCoul5}
\end{figure}

\begin{figure}
\noindent

\vspace{1cm}
\hspace{3.4cm}{\includegraphics[width=6.9cm]{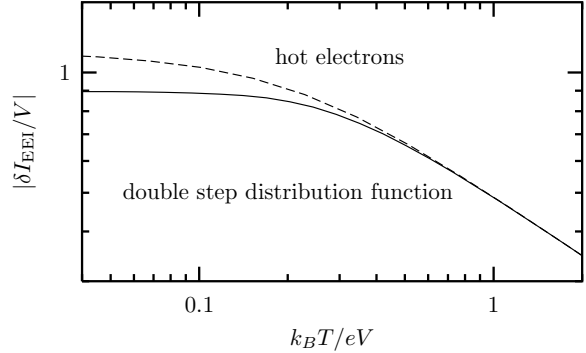}}
\caption{\label{figCoul6}Temperature dependence of the Coulomb interaction correction to the
conductance of a mesoscopic wire in arbitrary units. 
The full line corresponds to the double-step distribution function, the dashed line
corresponds to hot electrons with temperature profile, as explained in the text; $T$ is the electron
temperature in the reservoirs.}
\end{figure}

We now consider the other two  limits, b) and c).
We found that in both limits the current can be written as \cite{schwab01}
\begin{equation}
\delta I_{\rm EEI}(V,T) = {e^2 \over \hbar } {L_T \over L } f \left({eV / k_B T} \right) V
,\end{equation}
where the function $f$ depends on the distribution function and on the length of the wire; $T$ is the
temperature in the leads.
Numerical results are shown in Fig.~\ref{figCoul5}.
Notice that $f(eV/k_B T)$ is proportional to 
$\delta I_{\rm EEI}/V$, so that $f(eV/k_B T)$ also represents the voltage dependent
conductance in units of $(e^2/\hbar)(L_T/L)$.
For low voltage and large system size $L \gg L_T$, 
the result \cite{altshuler84} 
\begin{equation} 
\delta I_{\rm EEI}/V \approx -{e^2 \over \hbar }{ L_T \over L }\left[ {3 \zeta(3/2)\over (2\pi)^{3/2} } 
 - {5\over 6}{ L_T\over L} + {5 \sqrt{2} \zeta(5/2) \over \pi^{5/2} }{L_T^2 \over L^2}+  \dots  \right]; \quad
{ 3 \zeta(3/2) \over (2\pi)^{3/2} } \approx 0.498
\end{equation}
is approached. With decreasing length the correction to the current decreases, since 
electrons escape more quickly from the wire into the leads.
The full lines show the voltage dependent conductance for case c); the long-dashed line corresponds to case b).
The short-dashed line ($L/L_T=5$) is obtained within a simple approximation:
Instead of evaluating the full expression for $\delta {\bf j}_{\rm EEI}$
we take the linear conductivity as a function of temperature, and average over the temperature profile,
\begin{equation}\label{eqHeating}
\delta \sigma_{\rm heating} = {1\over L} \int_0^L \! \D x \, \delta \sigma ( T(x) )
.\end{equation}
Important results are the following:
in both cases b) and c) the conductance scales with voltage over temperature.
In case b) (hot electrons) the main effect is simple heating, {\it i.e.}~the non-ohmic effects 
are small.
In case c) (far from equilibrium) the current-voltage characteristic is quantitatively different 
from the hot electron regime. 
The temperature dependence of the Coulomb interaction contribution to the conductance 
is shown in Fig.~\ref{figCoul6} in a double logarithmic plot. 
The curves shown are obtained from the same numerical data for $f(eV/k_B T)$ as shown in Fig.~\ref{figCoul5} ($L/L_T=200$).
At high temperature the Coulomb correction to the conductance follows $1/\sqrt{T}$, which is seen as a
linear behavior in the double logarithmic plot.
When the temperature in the leads becomes lower than the voltage, the conductance saturates.
In the absence of inelastic scattering, c), this low temperature saturation appears at a higher temperature than
in b).

Finite size effects in the linear conductivity of thin wires of length $L \sim L_T$ have been studied {\em e.g.}\ by
Masden and Giordano \cite{masden82} who found a qualitative (although not quantitative) agreement
with the theoretical predictions.  
The distribution function in mesoscopic wires out of thermal 
equilibrium was measured by Pothier {\em et al.}~\cite{pothier97}.
For short wires and low temperature ($L\approx 1.5$ $\mu$m, $T\approx 25$ mK), 
the double step like distribution function was observed.
Unfortunately, as far as we know, there is no detailed investigation of the temperature and voltage 
dependence of the conductivity in this experiment. 
%
\subsection{Short wire with interfaces} 
\label{SecShortWire}
In subsection  \ref{SecBoundary} we demonstrated how to obtain the classical resistance of a
system which is composed of diffusive pieces and resistive interfaces in the
framework of the quasiclassical description.
In this subsection, following closely \cite{schwab02},
we include the Coulomb interaction, and in particular we will
concentrate
on structures of a size comparable to the thermal diffusion length $L_T$.
In subsection \ref{SecCoulInterface} we described a general 
formalism to obtain the quantum correction to the current through an
interface; here we apply the formalism to a specific geometry
where a diffusive piece of metal is connected via resistive interfaces to a
left  ($x=0$)
and right ($x=L$) reservoir.

We divide the quantum correction to the current into two contributions:
The first
has the meaning of a correction to the conductance, $\delta I^{(1)}= V  \delta G$.
The second 
can be interpreted as the 
redistribution of the voltage across the interfaces and the wire, $\delta I^{(2)} = G \delta V$.
More precisely we denote by $\delta I^{(2)}$ those terms which arise from the
correction to the distribution function like, for example,
\begin{equation}
\delta I^{(2)}_{\rm wire} = - e D N_0 {\cal A} \int \D \epsilon \, \partial_{\bf x} \delta F(\epsilon; {\bf x} )
;\end{equation}
we will give the explicit expression for $\delta I^{(1)}_{\rm wire}$ below.
Both terms are necessary in order to ensure current conservation.
In principle the variation of the distribution function, $\delta F(\epsilon; {\bf x} )$, can be
determined by solving the kinetic equation in the presence of interactions.
However by 
exploiting current conservation it appears that this is not necessary.
For the time independent problem which we consider here, current conservation means that 
the current does not depend on the position, 
{\em i.e}.\ 
$ I_l =  I_{\rm wire}= I_r$, where the subscripts $l$ and $r$ indicate the left and
right interface.
By fixing the voltage drop over the whole system to $V$, such that
the sum $\delta V_l + \delta V_r + \delta V_{\rm wire}$ is zero,  
it is possible to eliminate $\delta I^{(2)}$ from the equations to obtain 
the correction to the current as
\begin{equation} \label{eq162}
\delta I_{\rm EEI} = { R_l \delta I_l^{(1)} + R_r \delta I_r^{(1)} +
 R_{\rm wire } \delta I_{\rm wire }^{(1)} \over R_l + R_r + R_{\rm wire}}.
\end{equation}
To proceed further we need the explicit form of $\delta I^{(1)}_{l,r, {\rm wire}}$.
For an interface which is attached to an ideal lead on one side,
the quantum correction to the current is controlled by the correction to the
density of states on the other side,
\begin{eqnarray} \label{eq163}
\delta I_l^{(1)}  &= &- e {\cal A}\Gamma_l \int \D \epsilon  
  \delta N (\epsilon , 0) [ F(\epsilon; 0)- F^l(\epsilon)  ] \label{tunncurrleft}\\
\label{eq164}
\delta I_r^{(1)}  & = & - e  {\cal A} \Gamma_r \int  \D \epsilon 
  \delta N (\epsilon , L) [ F^r(\epsilon)- F(\epsilon; L) ].\label{tunncurrright}
\end{eqnarray}
Quite generally we obtain the density of states correction as 
\begin{eqnarray}
\delta N (\epsilon, { x} ) &= &
-N_0 \int { \D \omega \over 2 \pi}
S({ x},{ x}) \\
S({ x},{ x}) &= & - \Im  \int \D  { x}_1 \cr
 & & \times  F({\epsilon -\omega}; { x}_1) \rho_0( { x},\omega; { x}_1)
             \phi_{\rm ind}({ x}_1, \omega; { x})
,\end{eqnarray}
where $\rho_0( { x}, \omega;{ x}_1)$ describes the spreading of a charge injected into the system at 
${ x}_1$, and
$\phi_{\rm ind}$ is the induced electrical potential as we discussed in subsection \ref{SecInducedPotential}. 
By using the same notation for 
the correction to the current in a diffusive wire we find
\begin{eqnarray}
\label{wirecurr}
\delta I_{\rm wire}^{(1)}(x)  &=&  e D N_0 {\cal A} \int \D \epsilon \int {\D \omega \over 2 \pi}
\partial_x[ F(\epsilon; x) S(x, x)] \\
& &- 2 e D N_0 {\cal A} \int \D \epsilon \int {\D \omega \over 2 \pi} 
F(\epsilon; x) \partial_{x_1}S(x,x_1)|_{x_1=x}.
\nonumber
\end{eqnarray}
In general $\delta I_{\rm wire}^{(1)}(x)$ depends on the position $x$.
Equation (\ref{eq162}), however, is constructed in such a way that 
the spatial average $\delta I_{\rm wire}^{(1)} =L^{-1} \int \D x \delta I_{\rm wire}^{(1)}(x)$ has to be inserted.

We have already discussed the distribution function $F$ including the relevant boundary conditions 
in subsection \ref{SecBoundary}.
Inside a diffusive wire the charge density $\rho_0$ satisfies the diffusion equation, Eq.~(\ref{eq124}). 
At the boundaries to the left and right reservoir, one may derive the matching conditions
\begin{eqnarray}
D \partial_x \rho_0(\omega,x;x')|_{x=0} &= & \Gamma_l \rho_0(\omega,0;x') \cr
D \partial_x \rho_0(\omega,x;x')|_{x=L}  & = & - \Gamma_r \rho_0(\omega,
L;x')
,\end{eqnarray}
to be compared with the boundary conditions for the distribution function, Eqs.~(\ref{eq60}) and (\ref{eq61}).
A careful analysis is also required for the induced field $\phi_{\rm ind}(\omega, { x};{ x}')$.
We assume good metallic screening (compare subsection \ref{SecInducedPotential}) so that an injected charge is almost instantly screened,
and therefore the wire will be electrically neutral
with the possible exception of a thin surface layer. In this case 
inside the wire, one has
\begin{equation} \label{eqWireField}
-\sigma \partial_x^2 \phi_{\rm ind}(\omega, { x};{ x}') = e \delta({ x}-{ x}')
,\end{equation}
where $\sigma = 2 e^2 D N_0 {\cal A}$ is the conductivity.

From now on we concentrate on a short resistive wire with interfaces assuming 
temperatures of the order of and lower than the
Thouless energy, $\hbar D/L^2$. 
\begin{figure}
\centerline{ \includegraphics[height=4.9cm]{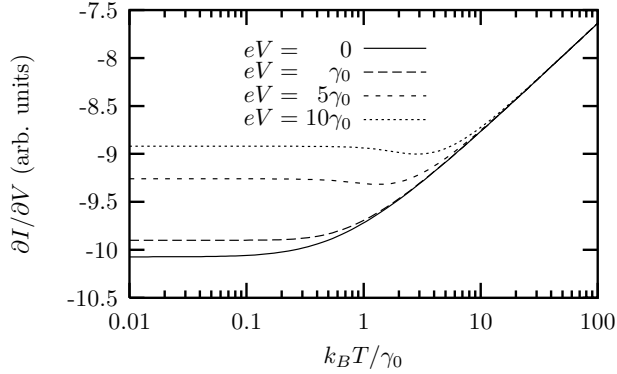} }
\caption{\label{Fig1}Contribution of the electron-electron interaction to the 
temperature dependence of the conductance of a wire with length $L \sim L_T$ 
for different values of the bias voltage, from \cite{schwab02}. 
The results are obtained under the assumption 
that the charge diffusion can be described with just a single diffusive
mode.
$\partial I/ \partial V$ is in units of $e^2/h$ and must be multiplied with
the non-universal number $A$ defined in the Eq.~(\ref{eq23}).
$\gamma_0$ is
the energy of the lowest diffusive mode; in the absence of interface barriers $\gamma_0 = \pi^2 \hbar D/L^2$,
for strongly resistive interfaces $\gamma_0 \to 0$.}
\end{figure}
We start by expanding the charge density $\rho_0$ in diffusive modes, 
\begin{eqnarray}
\label{modes}
\rho_0(\omega, x;x') &=  &e \sum_n { f_n(x) f_n(x') \over -\I \omega + \gamma_n}  
,\end{eqnarray}
where the (normalized) functions $f_n(x)$ are obtained from the eigenvalue equation
\begin{equation} \label{eq22}
-D \partial_x^2 f_n(x) = \gamma_n f_n(x)
.\end{equation}
In the zero-dimensional limit we approximate the sum in Eq.~(\ref{modes}) by retaining only
the eigenmode with the lowest energy, {\em i.e}.\
$\rho_\omega(x,x') \to e f_0(x) f_0(x')/(-\I \omega + \gamma_0)$.
This approximation is justified when the energy scales related to the 
temperature and to the voltage remain below the energy of the second lowest diffusive mode.
As a consequence of Eq.~(\ref{eqWireField}) the field $\phi_{\rm ind }(\omega, x;x')$ 
is frequency independent
and one observes that the frequency dependent factors in all the contributions
to the current are identical,
$\delta I(T,V) \sim  F^l_{\epsilon - \omega} F^r_{\epsilon}/(-\I \omega + \gamma_0)$.
The explicit result reads
\begin{eqnarray} \label{eq23}
\delta I_{\rm EEI} &= &
- A  {e \over 2 \pi } \int_0^\infty \! \! \D \eta \,  \E^{-\gamma_0 \eta}  
\left[{\pi T \over \sinh( \pi T \eta )} \right]^2 \! \sin( e V \eta ) 
,\end{eqnarray}
where only the dimensionless number $A$ and the quantity $\gamma_0$ depend on the 
details of the system under consideration.
A short time cut-off has to be introduced in the integral
in order to avoid a logarithmic divergence.
An explicit form can be obtained by refining the approximations made.
However we do not specify the cut-off here since it only leads to a temperature independent
correction to the linear conductance.

\begin{figure}
\centerline{ \includegraphics[height=4.9cm]{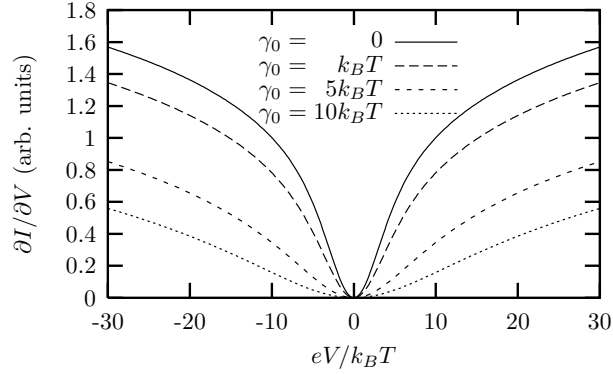} }
\caption{\label{Fig2}Voltage dependence of the conductance of a wire of length $L \sim L_T$. 
In all curves the linear conductance has been subtracted. The figure is taken from \cite{schwab02}.}
\end{figure}

Let us first  discuss the temperature and voltage dependence of $\delta I_{\rm EEI}$, 
then we determe $A$ and $\gamma_0$ explicitly in the two limits of
perfectly transparent interfaces and for interfaces with low transparency.
Figure \ref{Fig1} shows $\partial I /\partial V/(A e^2/2\pi)$ as a function of temperature,
a constant being subtracted.
At high temperature a logarithmic behavior is observed, 
\begin{equation}
{\partial I  \over  \partial V} = G(T) = G(T_0 ) + {e^2 \over 2 \pi \hbar} A  \ln(T/T_0)
,\end{equation}
which saturates below $T_{\rm sat} \sim \max(\gamma_0, eV)$.
Figure \ref{Fig2} shows the voltage dependence of the conductance; here the linear 
conductance has been subtracted.
For $\gamma_0 \ll T$ the conductance scales with voltage over temperature, while  when $\gamma_0$ is large the
relevant scale for conductance variations is $\gamma_0$.
Notice that the $\log T$ behavior and the current-voltage characteristic 
are universal for the zero-dimensional systems:
The results are not sensitive of whether the system is dominated by interfaces or diffusion,
nor to the shape of the sample, nor to the precise value of the conductance.
The microscopic details are only relevant via the number $A$ and the escape rate $\gamma_0$.

How large are the amplitude $A$ and the energy of the lowest diffusion mode $\gamma_0$?
In the case of two well transmitting interfaces, $G_{\rm wire} \ll G_l, G_r$, 
the eigenfunction of Eq.~(\ref{eq22}) with the lowest eigenvalue is
\begin{equation}
f_0(x) =  \sqrt{ 2 \over L } \sin( \pi x  /L), \gamma_0 = \pi^2 D/L^2. 
\end{equation}
The distribution function and the potential $\phi_{\rm ind}$ are determined as
\begin{eqnarray}
F(x) &  =  & [(L-x) F_l + x F_r] /L \\
\label{eq183a}
\phi_{\rm ind}( \omega, x; x') & = & {e \over G_{\rm wire} } 
\left\{ \begin{array}{lll}
(L-x')x/L^2 & & x< x' \\
(L-x) x'/L^2 & & x> x'
\end{array} \right. 
\end{eqnarray}
and the amplitude of the correction to the current is found to be $A =64/\pi^4 -4/\pi^2   \approx 0.25$.
In the opposite limit, $G_{\rm wire} \gg G_l, G_r$ the eigenvalue equation (\ref{eq22})
may be solved perturbatively in the barrier transparency and one obtains
\begin{eqnarray}
f_0(x) & = & 1/\sqrt{L}, \gamma_0 = (\Gamma_l + \Gamma_r )/L \\ 
F(x)  &  = & ( \Gamma_l F_l + \Gamma_r F_r )/(\Gamma_l + \Gamma_r) \label{eq29} \\
\phi_{\rm ind }( \omega, x;x')  & = & {e/( G_l + G_r) } \label{eq30}
,\end{eqnarray}
which leads to $A = 2 \Gamma_l \Gamma_r /( \Gamma_l + \Gamma_r )^2$.

It is useful at this point to  briefly discuss how the charging effects modify the above results.
We assume that the accumulated charge is proportional to the field as
$\rho(x)= \rho_0 + \rho_{\rm ind} = C \phi_{\rm ind}(x)$, where $C$ is a capacitance per unit length. 
From the continuity equation for the total charge density $\rho(x)$, one obtains a
diffusion equation for the field
\begin{equation}
-\I \omega C  \phi_{\rm ind} (\omega,x;x') - \sigma \partial_x^2 \phi_{\rm ind}(\omega, x;x') = e \delta(x-x')
.\end{equation}
For highly transmitting interfaces the solution is
\begin{equation}
\phi_{\rm ind}( \omega, x; x')
 =  2 e \sum_n {\sin(n \pi x / L) \sin( n \pi x'/L) \over
  - \I \omega (C L ) + (\pi n )^2 G_{\rm wire}}
,\end{equation}
and the correction to the current is modified according to
\begin{eqnarray}
\delta I_{\rm EEI}  &=  &- { e \over 2 \pi }
\int_0^\infty \D \eta \E^{-\eta \gamma_0 } 
\left[ { \pi  T \over \sinh ( \pi T \eta ) } \right]^2 \sin( e V \eta) \cr
&&\times \sum_n A_n\{1- \exp[-\eta (\pi n)^2/R(CL)] \}  \label{eq33}
.\end{eqnarray}
The numbers $A_n$ depend on the wave-function of the diffusive mode $n$, and $R=G_{\rm wire}^{-1}$.
Notice that although the free charge density $\rho_0(x)$ is zero-dimensional
many diffusive modes have to be taken into account in the field $\phi_{\rm ind}(x)$.
For a small capacitance $C$ the charging simply cures the short time divergence in the integral in Eq.~({\ref{eq23}}).
For a larger capacitance the $I$-$V$-characteristic is no longer a universal function.

For the system with poorly transmitting interfaces, the field in Eq.~(\ref{eq30}) 
has to be replaced by
$\phi_{\rm ind} = {e / (-\I \omega \tilde C + G_l + G_r ) }$, where we denote the capacitance of 
the system by $\tilde C$.
The modification to the current is analogous to Eq.~(\ref{eq33}).

Finally, as far as the experiments are concerned, a $\ln T$ behavior in the linear resistivity of 
short metallic bridges, together
with an $I$-$V$-characteristic which agrees well with the universal function 
(\ref{eq23}) has been observed in \cite{weber01}.
The bridges have been about 
$L=80$ nm long, $80$ nm wide and $10$ nm thick; the temperature region where the $\log T$ in
the linear resistivity has been seen is between $T\approx 100$ mK and $T\approx 2$ K.
In \cite{weber01} the Coulomb correction to the tunneling conductance has been suggested 
as the explanation;
the considerations presented here and in\cite{schwab02} 
show that, also in the intermediate regime with both diffusive and interface resistivity
the predicted $I$-$V$-characteristic does not change and thus agrees with the experimentally observed one.
In \cite{weber01} the Thouless energy, which sets the scale for the lowest diffusive mode in an 
open system and 
therefore for the low temperature saturation of the conductance,
has been estimated to be of the order of several Kelvin,
whereas the $\ln T$ is observed down to $100$ mK.
In the case with resistive interfaces, on the other hand, the energy of the lowest diffusive mode
is reduced,
$ \gamma_0  \sim  \hbar D/L^2 (R_{\rm wire} / R) \ll \hbar D/L^2$. 
Although some caution is necessary in applying these considerations to the 
experiment\footnote{The interface resistance required to obtain the low temperature saturation
of the conductance at $\sim $ 100 mK is about one order of magnitude
larger than the actually measured resistance; on the other hand there is still room to refine
the theory.},
we conclude that
in the measurement of \cite{weber01} the diffusive resistance may be considerably 
smaller than the interface resistance.
A further hint for the importance of interfaces is found from the prefactor $A$:
For the open system we found $A\approx 0.25$, and $A= 2 G_l G_r/(G_r+G_r)^2$ in the tunnel limit.
The experimental values \cite{weber01} are between $A\approx 0.43 \dots 0.7$, {\em i.e}.\ closer to the 
tunnel limit than to the open system. 
%
\subsection{Diffusive point contacts } 
\label{SecDiffDot}
In the previous subsection we observed that 
the electron escape rate $1/\tau_{\rm esc}=\gamma_0 $
determines the temperature below which the
Coulomb interaction contribution to the conductivity
saturates.
The reason for this saturation is the assumption that the 
system is connected, on both sides, to ``ideal leads'' 
where quantum fluctuations are completely suppressed.
In this subsection we will allow for voltage fluctuations in the leads, {\em i.e}.\ fluctuations
of the field $\check \phi_{\rm int}({\bf x},t)$,
which lead to a temperature dependent conductance also in the low temperature region
$k_B T < \hbar/\tau_{\rm esc}$.
When the system under consideration consists of a tunnel junction,
the fluctuations in the leads are dominated by the conventional 
Coulomb interaction effects on the tunnel conductance as we discussed in subsection \ref{SecCoulInterface}.

However, an interesting question is whether a tunnel junction is strictly necessary see Coulomb
blockade physics \cite{nazarov99,kamenev00,feigelman02, golubev01,levy01}.
As far as transport properties are concerned the issue was addressed by
Golubev and Zaikin \cite{golubev01}, and Levy Yeyati {\it et al.} \cite{levy01},
who 
studied within a scattering approach
the transport through a coherent scatterer coupled to an electromagnetic
environment.
For a single conducting channel,
the correction to the conductance due to the interaction with 
environmental modes was found to be ($T=0$)
\begin{equation}
{\delta G \over G  } =  - {2 e^2 \over h} ( 1-\tau ) \int_{eV} \D  \omega  {\Re Z \over \omega}
,\quad  G = {2 e^2 \over h}\tau, \end{equation} 
where $\tau$ is the channel transmission and the impedance $Z$ characterizes the environment.
For a poorly transmitting channel ($\tau \ll 1 $) this agrees with standard Coulomb blockade theory in the 
perturbative limit, 
compare 
Eqs.~(\ref{eq151a})--(\ref{eq155a}).
Apparently in a highly transmitting channel the relative change in the conductance is 
suppressed by the factor $(1-\tau)$ with respect to a tunnel junction.
In the multichannel case the reduction factor $(1-\tau)$ is replaced by \cite{golubev01}
\begin{equation}
\beta = {\sum_n  \tau_n (1- \tau_n ) \over \sum_n \tau_n  }
.\end{equation} 
The parameter $\beta$ is known from the theory of shot noise \cite{blanter00}; it is 
equal to one for tunnel junctions, and $\beta =1/3$ for diffusive conductors.

We will now derive the correction to the current through a diffusive conductor due to the interactions
in the leads within the quasiclassical theory.
When one neglects the Coulomb interaction inside the wire, the change of the Green's functions
due to the interaction in the leads fulfills the equations
\begin{eqnarray}
\left(  -\I \omega  + D \partial_{\bf x}^2 \right) \delta g^Z(\eta; {\bf x}, \omega ) &= &0 \\
\left(  -\I \omega  - D \partial_{\bf x}^2 \right) \delta g^K(\eta; {\bf x}, \omega ) &=  &0 
.\end{eqnarray}
The corresponding solutions decay or increase exponentially on the scale
$L_\omega = \sqrt{D/\omega}$. When the structure under consideration is longer than $L_\omega$,
the fluctuations in the leads influence $\delta g^{Z,K}$ only in a narrow region near the 
interfaces, and will only weakly affect the transport.
In the opposite limit, when the structure is shorter than $L_\omega$ or when considering
frequencies below the Thouless energy $D/L^2$, 
the approximate solutions of the equations are the linear interpolation of $g^Z$ and $g^K$
between the values on the left and right interfaces,
{\it i.e}.
\begin{eqnarray}
\delta g^Z(\eta;{\bf x}, t) & \approx  & {L-x \over L} \delta g^Z(\eta; 0, t)
                                        +  {x \over L} \delta g^Z(\eta; L, t) \label{195}\\
\delta g^K(\eta;{\bf x}, t) & \approx  & {L-x \over L} \delta g^K(\eta; 0, t) 
                                        + {x \over L} \delta g^Z(\eta; L, t)
.\label{196}\end{eqnarray}
One observes that the Coulomb correction to the current through the wire
can be expressed in terms of the correction to the Green's functions at the
interfaces between the wire and the leads.
In the following, for
simplicity, we consider interactions just in one of the two leads.
Under this condition we determine the correction to the current by inserting Eqs.~(\ref{195}) and (\ref{196})
into the general expression for diffusive systems obtained in Eq.~(\ref{103}), and integrate
between $0$ and $L$, with the result
\begin{equation}
\delta I_{\rm EEI} = - { 1\over 6 e }  { G } \int \D \epsilon {\delta N( \epsilon , 0) \over N_0  } 
                                        [ F(\epsilon, L) -  F (\epsilon, 0 ) ]
,\end{equation}
where $G$ is the conductance of the system  in the absence of fluctuations in the leads.
The result obtained within the quasiclassical method is consistent with
the scattering approach mentioned above: 
Similar to what is found for a tunnel junction, 
the Coulomb interaction correction to the conductance is related to the tunneling density of states
at the interfaces also for a fully coherent diffusive metal. For a diffusive conductor the variation of the 
conductance due to the interaction is reduced by a factor three compared to tunnel junctions.

The density of states at the interface is sensitive to the geometry of the leads.
When the leads have a  three-, two-, or one-dimensional character, the density of states in the leads
is proportional to
\begin{equation}
\delta N(\epsilon) \propto \left\{ 
\begin{array}{ll}
\sqrt{\epsilon}   & d=3\\
\ln (\epsilon )   & d=2\\
1/\sqrt{\epsilon} & d=1 
\end{array}
\right.
\end{equation}
at zero temperature.
Notice, however, that the density of states on the surface of a
three-, two-, or one-dimensional conductor is not identical to the density of states in the bulk.
For example, for a semi-infinite three-dimensional sample the density of states 
as a function of energy and of the distance from
a metal-insulator interface is \cite{altshuler84}
\begin{equation}
\delta N^{3d}(\epsilon, z) = {\sqrt{ \epsilon } \over \sqrt{2 } } {1 \over  2 \pi^2 D^{3/2}}
+ {1 \over 8 \pi^2 D z }
\left\{ 2 \int_{x_0}^{x_1} {\D x \over x} \E^{-x}\cos  x + \ln(\epsilon \tau ) \right\}
\end{equation}
with $x_0 = z \sqrt{2 \epsilon / D}$, $x_1 = z \sqrt{2/D\tau}$. The 
energy dependent part of the above equation varies thus by a factor two between the
surface and the bulk,
\begin{equation}
\delta N^{3d}_{\rm surface}( \epsilon )=  2 \delta N^{3d}_{\rm bulk}( \epsilon ) 
=  {\sqrt{ \epsilon } \over \sqrt{2 } } {1 \over  \pi^2 D^{3/2}} 
,\end{equation}
and the typical length scale for this variation is $L_\epsilon = \sqrt{D/\epsilon}$.
We expect that $ \delta N^{3d}_{\rm surface}( \epsilon )$ as it is given here is also applicable
to an interface between a three-dimensional lead and a diffusive point contact, provided
the diameter of the point contact is small compared to $L_\epsilon$.

In order to discuss the one-dimensional case we first go back to the general expression
for the density of states correction,
\begin{equation}
\delta N(\epsilon, {\bf x} ) = N_0 \Im \int {\D \omega  \over 2 \pi} \int \D{\bf x}_1 
F(\epsilon-\omega; {\bf x}_1 ) \rho_0({\bf x}, \omega; {\bf x}_1 ) \phi_{\rm ind}({\bf x}_1, \omega; {\bf x})
.\end{equation}
In metals the field $\phi_{\rm ind}$ usually propagates much faster than the charge density
$\rho_0$; in particular, in the limit of good screening, the charge neutrality condition  
corresponds to a field $\phi_{\rm ind}$ which propagates instantly over
the full system under consideration.
As a consequence the field $\phi_{\rm ind}({\bf x}_1, \omega; {\bf x}) $ is in many situations a smooth
function of the position, compared to the charge density $ \rho_0({\bf x}, \omega; {\bf x}_1 )$ which is 
concentrated in a region $ |{\bf x}-{\bf x}_1| \sim \sqrt{D/\omega}$.
In one dimension, for example, the induced potential varies linearly in the position with
the typical scale of the system size, see for example Eq.\ (\ref{eq183a}).
When neglecting the spatial variation of both $\phi_{\rm ind}$ 
and the distribution function, then 
the variable ${\bf x}_1$ in the equation above can easily be integrated to yield 
the density of states as
\begin{eqnarray}
\delta N(\epsilon, {\bf x} ) &\approx & N_0 \Im \int {\D \omega  \over 2 \pi}  
F(\epsilon-\omega; {\bf x} ) {e \over -\I \omega} \phi_{\rm ind}({\bf x}, \omega; {\bf x}) 
.\end{eqnarray}
When the leads are in thermal equilibrium the first and second derivative of the current with respect to the voltage
are ($T=0$)
\begin{equation}
{\partial I\over \partial V }  = G + {1 \over 3} G {\delta N(eV, 0 ) \over N_0 },
\end{equation}
and
\begin{equation}
{\partial^2  I \over \partial V^2 }  = {2 \over 3} { G \over V } { \Re Z(eV,0) \over h/e^2 }
.\end{equation}
In the last equation we replaced the induced electrical potential by
the local impedance, $\phi_{\rm ind}({\bf x}, \omega ; {\bf x}) = e Z(\omega, {\bf x})$.
Remember that $\phi_{\rm ind}({\bf x}, t; {\bf x})$ is the voltage at $({ \bf x} , t)$, due to a 
charge that has been placed at ${\bf x}$ and $t=0$. 
At the interface between a point contact and leads the voltage decays due to the charge flowing into the leads or into
the contact, and therefore $Z(\omega, {\bf x}=0)$ is the total impedance of the leads and the point contact in parallel.
For example, when we describe the point contact by a capacitance $C$ and a resistance $R$,
and the leads as a $RC$-transmission line with a resistance $R_0$ and a 
capacitance $C_0$ per unit length, the impedance $Z$ is given by \cite{nazarov89c}
\begin{eqnarray}
Z^{-1} & =  &Z_{\rm lead}^{-1} +{Z_{\rm point \, contact}^{-1} }  \\
Z_{\rm point\, contact}^{-1} & = & -\I \omega C + R^{-1} \\
Z_{\rm lead} & = & \left\{      
\begin{array}{ll} 
R_0 L                        & {\rm if}  \,  L <     \sqrt{1/ \omega C_0  R_0  } \\[1mm] 
\sqrt{R_0/- \I \omega C_0}           & {\rm if}  \,  L > \sqrt{1/ \omega C_0  R_0  } 
\end{array}
\right. 
. \end{eqnarray}
For long leads, where
$Z(\omega)$ is proportional to $1/\sqrt{\omega}$,
the density of states correction acquires its usual one-dimensional form, 
$\delta N(\epsilon  ) \propto 1/\sqrt{\epsilon} $.

The situation where $Z_{\rm lead}(\omega) $ is frequency independent has been analyzed in detail by Golubev and
Zaikin in \cite{golubev01},
with the result
\begin{equation}
{\partial I\over \partial V }  = {1\over R} - {1\over 3}{R_{\rm tot} \over R }{e^2 \over h } 
\left\{
\begin{array}{ll}
\ln \left[ 1 + (  \hbar  /  eV   R_{\rm tot} C )^2   \right] & {\rm for } \, T  \to 0   \\[1mm]
2 \gamma + 2+ 2 \ln\left( {\hbar/  R_{\rm tot} C  \over  2 \pi k_B T  } \right)   &  {\rm for } \, V  \to 0
 \end{array}
\right.
,\end{equation}
with $R_{\rm tot}^{-1} =  R^{-1}+ (R_0 L)^{-1}$, and $\gamma \approx 0.577$.

Golubev and Zaikin suggested that the $\ln T $ behavior in the conductance, and the
zero bias anomaly observed by Weber {\em et al}.\ \cite{weber01} 
in short metallic bridges, might be due to the effects described above;
in this case one concludes that the electromagnetic environment of the bridges
behaves as an Ohmic resistor, with the environmental resistance at least comparable to or larger than the resistance
of the nanobridge ($R \sim  10 \; \Omega$);
for related experiments see also \cite{yu03}.

In concluding this section we summarize the basic mechanisms by which the Coulomb interaction
affects transport through small structures. 
At low temperatures -- in the absence of inelastic scattering --  
the interaction effects may be grouped in two classes.
First, the internal electric potential created in a diffusive conductor leads to an additional source
of random scattering that interferes with scattering from impurities. This is a quantum effect and depends
on the details of charge diffusion. We have discussed in subsections \ref{SecLongWire} and
\ref{SecShortWire} how charge diffusion is modified in small
structures, and how this in turn modifies transport. 
Second, adding a charge to a conductor costs energy and therefore the transport is blocked.
This is avoided in macroscopic samples via screening,
but the latter can be inefficient in small structures, where the necessary flow of the background charge
is slow due to the geometric constraints.
We have discussed this type of physics in 
subsections \ref{SecCoulInterface} and \ref{SecDiffDot}, where
the screening properties were taken into account in terms of an effective environmental impedance $Z$.

Of course, in real systems, the two groups of interaction effects may be difficult to separate. 
However, in a large
number of experimentally relevant cases, this separation is indeed possible due to the different time scales
for screening and for charge diffusion. In ordinary metals, for instance, screening is 
much faster than charge diffusion. 
%
\section{Summary}\label{SecSummary} 
In this article we discussed transport properties of mesoscopic samples
with emphasis on electron-electron interaction effects. 
The theory was formulated in terms of the quasiclassical Green's functions. These Green's functions 
are averaged over impurities from the beginning, and therefore the complex diagrammatic
analysis is avoided.
We demonstrated that many known effects can be obtained within this formalism:
the interaction corrections to the diffusive transport, density of states, and also 
non-perturbative Coulomb blockade physics.

A major part of this article has been devoted to the regime of diffusive transport, where 
we worked out a theory for electron-electron interaction effects far from thermal equilibrium. 
There is a long list of possible extensions:
for example it would be interesting to study time dependent phenomena.
Response to time dependent fields with frequencies in the microwave regime has been studied in some
detail for weak localization, both theoretically \cite{altshuler85,kravtsov93,volker02} and 
experimentally \cite{vitkalov88,giordano91}. 
In these experiments the influence of a microwave field on the DC-conductivity of thin films was investigated.
In the experimental studies, besides weak localization, the
interaction effect also is present, but
there are only few theoretical studies of the interaction correction in the presence of high frequency fields.
Altshuler {\it et al.}~\cite{altshuler85} calculated the high-frequency (linear) conductivity.
A first attempt to calculate the DC-conductivity in the presence of microwaves
was made by Raimondi {\it et al.}~\cite{raimondi99}.
There the microwave field was included in the theory via the diffusion propagators. The distribution
function, on the other hand, was considered stationary.
Going beyond this analysis, the distribution function could be determined by explicitly solving
the kinetic equation in the presence of the microwave. 
The thus determined distribution function has then to be used in 
the expression for the interaction correction to the current density, Eq.~(\ref{eqCorr46}).

In Section \ref{SecNanNo}
we discussed transport properties of simple devices such as films and wires with interfaces. 
This immediately suggests that our formalism 
may be applied to hybrid systems,
where also interfaces between different 
materials are present, {\em e.g}.\ metal-superconductor, metal-ferromagnet, etc.
Furthermore we restricted ourselves to the limit of diffusive electron motion; in small and clean samples, however,
the mean free path can be longer than the systems size, which requires a theory 
for ballistic electron motion.
A theory for the Coulomb correction to transport in ballistic systems has recently been put forward
by Zala {\em et al.}~\cite{zala01} in order to describe transport at intermediate temperatures, where
the thermal time $\tau_T$ is comparable to the (elastic) scattering time $\tau$. 
At present, however, the theory has not yet been applied to finite size systems.

We conclude the list of possible extensions of the work presented here by 
noting that we have
concentrated on the ensemble averaged current density, leaving aside the rapidly developing field  
of current noise \cite{blanter00,gutman01,noiseCounting}.

In relation to the problem of electron dephasing, we have shown that
dephasing affects also the interaction correction to the conductivity. 
However there are still many open questions:
What is the microscopic nature of the low temperature saturation of the dephasing time?
What is the microscopic origin of the dephasing rate which varies as $\sqrt{T}$ in gold films \cite{mohanty98},
and as $T^{1/3}$ in $\rm Bi_2Sr_2CuO_6$ \cite{jing91}?
Why is the dephasing rate, as determined from the weak localization magneto-resistance, different
from the rate determined from conductance fluctuations measurements \cite{hoadley99,aleiner02b}?

To summarize, transport in mesoscopic- and nano-structures is an active field of research, with 
several open questions ranging from fundamental problems of quantum theory to possible technological
application in the future.

\begin{acknowledgement}
We acknowledge discussions with C.\ Castellani, T.\ L\"uck, and U.\ Eckern. 
This work was supported by the German Italian exchange program Vigoni (CRUI and DAAD) and MIUR-COFIN2002022534.
\end{acknowledgement}

\appendix
\section{Field theoretic approach: the non-linear sigma model}
\label{SecAppA}
The field theoretic formulation of the interacting, disordered electron
system was pioneered by Finkelstein in the 1980's \cite{finkelstein83}.
Recently, this formulation has been extended to the non-equilibrium case by means of the Keldysh technique
by Kamenev and Andreev \cite{kamenev99}, and Chamon, Ludwig and Nayak
\cite{chamon99} in the case of normal-conducting metals, and by
Feigel'man, Larkin and Skvortsov \cite{feigelman00} in the case of superconductors.
Given the already extensive literature available on the subject, we
believe that, rather than repeating the derivation of the
non-linear sigma model, it is perhaps more useful,
to show  how 
the formalism applied in the present paper -- namely quasiclassical Green's function
technique in the presence of fluctuating internal fields -- is found when starting from the
non-linear sigma model.

In the absence of the electron interaction, the action of the non-linear sigma model is
given by
\begin{equation}
\I S_0 = -{\pi N_0 \over 4}
\left[ D \Tr (\partial_{\bf x} Q)^2 + 4 
\Tr \partial_t Q  \right],
\end{equation}
where the trace is a summation or integration over all 
degrees of freedom, and the time derivative is here understood to be
$\partial_t Q = \partial_t Q_{t t_1}|_{ t_1 = t}$.
The field $Q\equiv Q^{ij}_{tt'}({\bf x})$ must satisfy the  constraint
 $Q^2=1$, {\em i.e}.\
\begin{equation}
\sum_j \int \D t \,  Q_{t t_1 }^{ij}({\bf x}) Q_{t_1 t'}^{jk}({\bf x}) = \delta(t-t') \delta_{ik}
.\end{equation}
An external scalar field $\phi^{\rm ext}$ and the Hubbard-Stratonovich field $\Phi$
due to the Coulomb interaction are described by the 
following terms in the action:
\begin{eqnarray}
\I S_1 &= & -\I \pi N_0 \Tr \Phi_\alpha \gamma^\alpha Q
+ \Tr \Phi_{\alpha} V^{-1} \sigma^x_{\alpha \beta} \Phi_{\beta} \\
\I S_2 &= & - \I \pi N_0 \Tr  e \phi^{\rm ext} Q
+ 4 N_0 \Tr e\phi^{\rm ext} \Phi_2
,\end{eqnarray}
with $\gamma^1 = \sigma^0$ (the $2\times2$ unit matrix) and $\gamma^2 = \sigma^x$, and $V$ is the 
``statically screened'' interaction, {\em i.e}.\ a constant.

In order to make contact with the quasiclassical approach, we express the
Green functions in terms of the  $Q$-fields.
First we observe that \cite{kamenev99}
\begin{equation}
\check G_{tt'}({\bf x}, {\bf x'} )=
\langle \left[G_0^{-1} +{ \I \over 2 \tau} Q +
\Phi_\alpha \gamma^\alpha  \right]^{-1}  \rangle_{Q,\Phi}
,\end{equation}
where the brackets $\langle \dots \rangle_{Q,\Phi}$
indicate that one has to average over the fields $Q$ and $\Phi$.
By using the condition that the fields $Q$ and $\Phi$ are only slowly
varying in space and time, one finds a relation between the $\xi$-integrated
Green function and the $Q$-matrix.
In particular, upon taking the  $s$-wave component one finds
\begin{equation}
 \int {\D \hat {\bf p }\over 4 \pi } \check g_{tt'} ({\bf {\hat p}}, {\bf x} ) =
 \langle  Q_{tt'}({\bf x}) \rangle_{Q,\Phi}
.\end{equation}

At this point we make the crucial approximations:
At first we restrict the field $Q$ to a $\Phi$-dependent saddle point configuration.
Varying the action with respect to $Q$, under the condition $Q^2=1$, we find
\begin{equation}
{1\over 2} \pi N_0 D \partial_{\bf x}^2 Q - \pi N_0 \partial_t \delta(t-t_1) \delta ({\bf x} - {\bf x}_1 )  
-\I  \pi N_0 ( e \phi^{\rm ext} + \Phi_\alpha \gamma^\alpha )
+ \lambda Q + Q \lambda =0
.\end{equation}
where $\lambda \equiv  \lambda^{ij}_{tt_1}({\bf x}) $ is a Lagrange multiplier.
Making use of the normalization condition and after eliminating $\lambda$
the saddle point equation reads
\begin{equation}
\partial_t Q - D \partial_{\bf x } ( Q \partial_{\bf x}  Q )
-\I \left[e \phi^{\rm ext} + \Phi_\alpha \gamma^\alpha, Q  \right] = 0
,\end{equation}
which agrees with the kinetic equation for the quasiclassical
Green's function in the presence of an internal field. Hence 
we identify $Q$ with $g$.
In the next step we restrict $\Phi$ to a saddle point plus quadratic fluctuations.
Variation of $S( Q[\Phi], \Phi )$ with
respect to $\Phi$ gives the saddle point
for the internal field,
\begin{eqnarray}
\Phi_1^{\rm MF}({\bf x}, t)  
 & = & V \pi N_0 \left[Q^{12}_{tt}( {\bf x} )  - {2 \over \pi } e \phi^{\rm ext}({\bf x}, t )  \right] \\
\Phi_2^{\rm MF}({\bf x}, t ) &=& 0
\end{eqnarray}
In order to obtain this relation we used that $Q^{21}=0$, and $Q^{11}_{tt}+ Q^{22}_{tt} =0$
at the saddle point.
The field fluctuations are determined from the second derivative of the action with respect 
$\Phi$,
\begin{equation}
{\D^2 S \over \D \Phi^2 } =
\left( {\D \over \D \Phi } {\delta S \over \delta Q } \right) {\delta Q \over \delta \Phi }
+ { \delta S \over \delta Q   } { \delta^2 Q \over \delta \Phi^2 }
+ {\D \over \D \Phi } {\delta S \over \delta \Phi }
.\end{equation} 
Since ${ \delta S / \delta Q   }=0 $ for all field configurations $\Phi$ the first two
terms in the equation above are equal to zero.
The final result is 
\begin{equation}
-\I \langle \Phi_i({\bf x}, t) \Phi_j({\bf x}', t' )  \rangle = {1 \over 2} V^{\rm dyn}_{ij} ({\bf x}, t ; {\bf x}', t' )
,\end{equation}
where the dynamically screened interaction $V^{\rm dyn}$ is 
\begin{equation}
\left[ V^{\rm dyn} \right]^{-1} =
V^{-1}\delta({\bf x} - {\bf x }') \delta (t-t') 
\left( \begin{array}{cc}
0 & 1 \\
1 & 0
\end{array} \right)
+ 
\left( \begin{array}{cc}
0 & \chi^R({\bf x}, t; {\bf x}', t')\\
\chi^A({\bf x}, t; {\bf x}', t')    & \chi^K({\bf x}, t; {\bf x}', t')
\end{array} \right)
\end{equation}
with
\begin{eqnarray}
\chi^R({\bf x}, t; {\bf x}', t') &= & -\pi N_0 
{ \delta Q^{12}_{tt}({\bf x}) \over \delta \phi_1({\bf x}', t')} = \chi^A( {\bf x}', t'; {\bf x}, t)  \\
\chi^K({\bf x}, t; {\bf x}', t') &= & 
-\pi N_0 {\delta Q^{12}_{tt}({\bf x})  \over \delta \phi_2({\bf x}', t') }
-\pi N_0 {\delta Q^{21}_{tt}({\bf x})  \over \delta \phi_2({\bf x}', t') }
.\end{eqnarray}

\section{Fermi liquid parameters and the spin triplet channels}
\label{SecAppB}
In the entire article we took into account the electron spin
only by factors of two appearing at various places.
When considering the electron spin explicitly,
the Green's function acquires  spin indices, and the Zeeman term has to be taken into account
in the  equation of motion:  
\begin{equation}
\partial_t \check g - D \partial_{\bf x} (\check g  \partial_{\bf x}  \check g ) 
- \I [ e \check \phi +\mu_B \boldsymbol\sigma \cdot  \check {\bf B} , \check g ] = 0
.\end{equation}
Both the scalar field $\check \phi$ 
and the magnetic field $\check {\bf B}$ are a sum
of internal and external contributions, 
$ \check \phi    = \phi^{\rm ext}    \, \check 1 + \check \phi^{\rm int} $,
$ \check {\bf B} = {\bf B}^{\rm ext} \, \check 1 + \check { \bf B}^{\rm int}$.
By separating the spin singlet and triplet components of the Green's function as
\begin{equation}
g_{ss'} =  g \delta_{ss'} + {\bf g} \cdot {\boldsymbol \sigma_{ss'} }, \quad
g   = {1\over 2} \sum_{s} g_{ss}, \quad
{\bf g}  =  {1\over 2} \sum_{ss' } {\boldsymbol \sigma_{ss'} } g_{s's}
,\end{equation}
the charge density and magnetization density are expressed as 
\begin{eqnarray}
\rho({\bf x}, t)    & = & 2     e N_0 \left[ {\pi \over 2} g^{12}({\bf x}, t) - e \phi_1({\bf x}, t) \right] \\
{\bf m}({\bf x}, t) & = & 2 \mu_B N_0 \left[ {\pi \over 2} {\bf g}^{12}({\bf x}, t) - \mu_B {\bf B}_1({\bf x}, t) \right] 
,\end{eqnarray}
where $\phi_1$, ${\bf B}_1$ are the sum of the external field and the diagonal component of 
the corresponding internal field.
As in the spinless case we assume that 
the internal field has Gaussian fluctuations around some mean value.
The mean field equations read
\begin{eqnarray}
\label{meanfield1}
\phi_1^{\rm MF}    & = & A_0^s \left[ {\pi \over 2 }{1\over e} g^{12}_{tt}({\bf x})
 - \phi^{\rm ext} \right], \\
 \label{meanfield2}
{\bf B}_1^{\rm MF} & = & A_0^a \left[ {\pi \over 2 }{1 \over \mu_B } 
{\bf g}^{12}_{tt}({\bf x})  - {\bf B}^{\rm ext} \right].
\end{eqnarray}
On average, the quantum components of the fields are zero, $\phi_2^{\rm MF}= {\bf B}_2^{\rm MF} =0$.
In the equations above $A_0^s$ and $A_0^a$ are the Landau parameters, which have the following relations
to the standard parameters $F_0^{a,s}$, and to the interaction constants in the spin singlet and triplet channels,
$\gamma_{s,t}$:
\begin{eqnarray} 
 A_0^a & =  & {F_0^a \over 1 + F^0_a }  = 2 \gamma_{t} \\
 A_0^s & =  & {F_0^s \over 1 + F^0_s }  = 2 \gamma_{s}
.\end{eqnarray}
Before discussing the fluctuations we demonstrate how to obtain 
the charge susceptibility
within the mean field scheme. The variation of the charge density in the presence of
an applied external field
$\phi^{\rm ext}$ is
\begin{eqnarray}
\delta \rho({\bf x}, t)& = &  2 e N_0 \left[ {\pi \over 2} 
\delta g^{12}_{tt}({\bf x}) - e (1-A_0^s) \phi^{\rm ext}({\bf x}, t) 
-A_0^s {\pi \over 2} \delta g^{12}_{tt}({\bf x}) \right] \\
 &= &2 e N_0 (1-A_0^s) \left[ {\pi \over 2} \delta g^{12}_{tt}({\bf x})  - e\phi^{\rm ext}({\bf x}, t)  \right]
;\end{eqnarray}
here we made use of the self-consistency condition for the internal field,
Eqs.~(\ref{meanfield1}) and (\ref{meanfield2}).
The variation of the Keldysh component of the Green's function due to the applied field
is determined by the equation of motion.
At first we note that the commutator of the field with the Keldysh Green's function at equal times is 
\begin{equation} 
\left( [  \phi, g^{12} ] \right)_{tt}= - {2 \I \over \pi } \partial_t \phi(t) . 
\end{equation}
By using this relation it is found that
\begin{equation} \label{eq241}
( - \I \omega + D q^2 ) \delta g^{12}({\bf q}, \omega) = {2 \over \pi} (-\I \omega ) 
           \left[ e \phi^{\rm ext}({\bf q}, \omega) + e \phi^{\rm MF}_1({\bf q}, \omega) \right]
, \end{equation}
and then
\begin{equation}
\left[ - \I \omega (1-A_0^s) + D q^2 \right] \delta g^{12}({\bf q}, \omega) = {2 \over \pi} (-\I \omega )(1-A_0^s) 
           e \phi^{\rm ext}({\bf q}, \omega)
.\end{equation}
Finally, we find the charge susceptibility in the presence of 
Fermi liquid renormalizations,
\begin{equation}
\chi({\bf q}, \omega ) = 2 e^2 N_0 (1-A_0^s) {Dq^2 \over -\I \omega (1-A_0^s) + D q^2}
.\end{equation}
In particular the static susceptibility is modified by the factor $(1-A_0^s)$ in the
presence of the interactions.
Similar results are obtained for the spin susceptibility.
The susceptibility, obtained here using the quasiclassical formalism, agrees with 
the direct diagrammatic approach, see {\em e.g}.~\cite{castellani84}.

 The field fluctuations are also fully 
determined within our formalism, as we discussed in the previous section.
Here we give the expression in the absence 
of an external magnetic field, and for a spatially homogeneous system.
The retarded interaction in the spin singlet channel is 
\begin{equation} \label{eq244}
-\I e^2\left\langle \delta \phi_1  \delta \phi_2 \right\rangle = { A_0^s \over N_0 } {- \I \omega  +  D q^2 \over -\I (1-A_0^s) \omega + Dq^2 } 
\end{equation}
and in the spin triplet channel,
\begin{equation} \label{eq245}
-\I \mu_B^2 \langle \delta B_1^i  \delta B_2^j \rangle
= { A_0^a \over N_0 } {- \I \omega  +  D q^2 \over -\I (1-A_0^a) \omega + Dq^2 } \delta_{ij}
,\quad i,j=x,x,z .\end{equation}

%
%


\end{document}